\documentclass[nofootinbib,superscriptaddress,showkeys]{revtex4-1}

\usepackage{amsmath,amssymb,amsthm}

\usepackage{hyperref}
\usepackage{amsthm}
\usepackage{graphicx}
\usepackage{subfigure}
\usepackage{color,bbm}
\usepackage[normalem]{ulem}
\usepackage{color}
\usepackage{gensymb}

\definecolor{orange}{rgb}{1.00,0.50,0.0}

\renewcommand{\S}{\mathbbm{S}}
\newcommand{\R}{\mathbbm{R}}

\newtheorem{theorem}{Theorem}

\newtheorem{definition}[theorem]{Definition}
\newtheorem{proposition}[theorem]{Proposition}

\begin{document}

\title{Parsimony, exhaustivity and balanced detection in neocortex}

\author{Alberto Romagnoni\footnote{${}^{\! \! \mathrm{,b}}$ These authors contributed equally.}${}^{,}$}
\email{alberto.romagnoni@college-de-france.fr}

\affiliation{Mathematical Neuroscience Team, CIRB - Coll\`ege de France (CNRS UMR 7241, INSERM U1050, UPMC ED 158, MEMOLIFE PSL), 11 Place Marcelin Berthelot, 75005 Paris}
\author{J\'er\^ome Ribot${}^{\mathrm{a,}}$}
\affiliation{Mathematical Neuroscience Team, CIRB - Coll\`ege de France (CNRS UMR 7241, INSERM U1050, UPMC ED 158, MEMOLIFE PSL), 11 Place Marcelin Berthelot, 75005 Paris}
\author{Daniel Bennequin${}^{\mathrm{b,}}$}
\affiliation{G\'eom\'etrie et dynamique, Universit\'e Paris Diderot (Paris VII), Paris, France}
\author{Jonathan Touboul${}^{\mathrm{b,}}$}
\affiliation{Mathematical Neuroscience Team, CIRB - Coll\`ege de France (CNRS UMR 7241, INSERM U1050, UPMC ED 158, MEMOLIFE PSL), 11 Place Marcelin Berthelot, 75005 Paris}
\affiliation{INRIA Mycenae Team, Paris-Rocquencourt, France.}

\date{\today}%

\begin{abstract}
One fascinating aspect of the brain is its ability to process information in a fast and reliable manner. The functional architecture is thought to play a central role in this task, by encoding efficiently complex stimuli and facilitating higher level processing. In the early visual cortex of higher mammals, information is processed within functional maps whose layout is thought to underlie visual perception. The possible principles underlying the topology of the different maps, as well as the role of a specific functional architecture on information processing, is however poorly understood. We demonstrate mathematically here that two natural principles, local exhaustivity of representation and parsimony, would constrain the orientation and spatial frequency maps to display co-located singularities around which the orientation is organized as a pinwheel and spatial frequency as a dipole. This observation is perfectly in line with new optical imaging data on the cat visual cortex we analyze in a companion paper. Here we further focus on the theoretical implications of this structure. Using a computational model, we show that this architecture allows a trade-off in the local perception of orientation and spatial frequency, but this would occur for sharper selectivity than the tuning width reported in the literature. We therefore re-examined physiological data and show that indeed the spatial frequency selectivity substantially sharpens near maps singularities, bringing to the prediction that the system tends to optimize balanced detection between different attributes. These results shed new light on the principles at play in the emergence of functional architecture of cortical maps, as well as their potential role in processing information.
\end{abstract}


% \abbreviations{SF, spatial frequency; OR, orientation; PC, Pinwheel center}

\maketitle

In the neocortex, the part of the mammalian brain in charge of higher functions, multiple sensory modalities are represented. Characterizing finely these functional and anatomical organizations has been a great success of the past decades, in part thanks to great advances in cortical imaging techniques, and we now dispose of a relative clear description of the neocortex architecture. However, the principles that govern these architectures, as well as their role in efficiently encoding and decoding information, remain largely unknown, and are central concepts for comprehending how the brain perceives and processes information.

The early visual cortex of higher mammals provides a particularly interesting framework since it contains the concurrent representation of multiple attributes of the visual scene, processed into parallel cortical maps whose layouts is commonly thought to be mutually interdependent. Groups of neurons in this area are preferentially selective to one specific value for each attribute. For instance, in response to a drifting grating, neurons in the primary visual area V1 encode the orientation (OR)~\cite{hubel1959receptive} of the stimulus as well as its spatial frequency (SF)~\cite{movshon1978spatial}. 
The two-dimensional orientation map is continuous and consists of regular domains where preferred OR varies smoothly together with singularities, the pinwheel centers (PC), around which all orientations are represented~\cite{bonhoeffer1991iso,bonhoeffer1995optical}. Moreover, the local nature of the visual scene is retinotopically encoded in the primary visual area~\cite{hubel1959receptive}: the information of a specific zone of the visual scene is processed by nearby neurons~\cite{tusa1978retinotopic}, and brain areas organizing these neurons reproduce the same characteristics at several places into a quasi-periodic structure~\cite{kaschube2010universality}.

Within a fundamental domain around a PC, the orientation map is locally exhaustive (all attributes are represented), yet it is parsimonious in the sense that any orientation is represented along a single level set. These two principles constitute very natural candidates for organizing the maps, yielding specific zones receiving all the information of the visual scene in an economic manner. The study of the representation of other attributes may allow investigating whether these principles also constrain their layout. 

Among other possible functional organization in the visual cortex, the spatial frequency has recently attracted much interest. A common view is that its organization is constrained to that of the orientation in order to ensure a uniform coverage, i.e. an even representation of the pairs (OR,SF). This theory was supported by data reporting an orthogonal relationship between iso-SF and iso-OR lines~\cite{nauhaus2012orthogonal,yu2005coordinated} or the fact that PCs shall be situated near extrema of the SF representation~\cite{shoham1997spatio,issa2000spatial,issa2008models}. These evidences did not appear clearly across different species: while strong orthogonality has been reported at global scale in monkey~\cite{nauhaus2012orthogonal}, only a weak tendency to orthogonality was shown in ferret~\cite{yu2005coordinated}. In cat, it remains a disputed issue. Indeed, it was recently shown that the distribution of angles between iso-OR and iso-SF lines were not peaked around $90\degree$: these are globally uniform, with a small bias towards alignment in the vicinity of PCs~\cite{ribot2013organization}. 

This context motivated us to come back to this problem using new optical imaging recordings of the SF preference in V1 as well as mathematical modeling. We mathematically demonstrate here that SF representations that satisfy our two candidate principles, namely that are locally exhaustive and optimally parsimonious, organize around singular points into a universal topology evocative of an electric dipole potential. Using our new high resolution optical imaging data in cat, we show in a companion paper~\cite{ribot14} that indeed, the SF map is continuous with dipolar singularities co-localized with the PCs. In the present paper, we rather focus on the consequences of such an architecture on the coding capabilities. We show using a computational model that pinwheel-dipole (PD) architectures, even if they do not allow even representation of (OR,SF), may improve perceptual precision compared to the orthogonal architecture. Going deeper in the coding capabilities of PD architectures, we realize that these organizations leave room for balanced detection of both attributes, but this occurs for SF selectivities sharper than the value previously reported in the literature~\cite{ribot2013organization}. Using finer estimates of the selectivity in the vicinity of PCs, we show indeed a clear sharpening of the SF selectivity near PCs perfectly consistent with the computational value predicted by the trade-off, leading to the natural prediction that PCs are singular locations of several maps at which selectivity ensures balanced detection. 

\section{Results}
\subsection{Universal topologies of minimal redundancy maps}\label{sec:topology}
Finding an optimal topology satisfying few simple conditions is easier said than done. A striking feature of the OR map is its very specific organization around singularities, the pinwheel topology, where the map is locally exhaustive and parsimonious. We will show that these two principles characterize univocally the topology of maps representing periodic (e.g, OR) or non-periodic quantities (e.g, SF). 

Because of the quasi-periodic structure of visual representations and the local nature of our criteria, we restrict our analysis to a small region of the visual cortex defined by an open set $\Omega$, which is assumed to be, without loss of generality, a disc. The orientation map is therefore defined as a continuous function $f:\Omega\mapsto \S^1$ where $\S^1$ is the disc $[0,\pi]$ where we identify $0$ and $\pi$. The SF map $g$ is also defined on $\Omega$, and takes values on an open (non-periodic) interval $U\subset \R$.

\begin{figure}[!h]
	\centering
		\includegraphics[width=.9\textwidth]{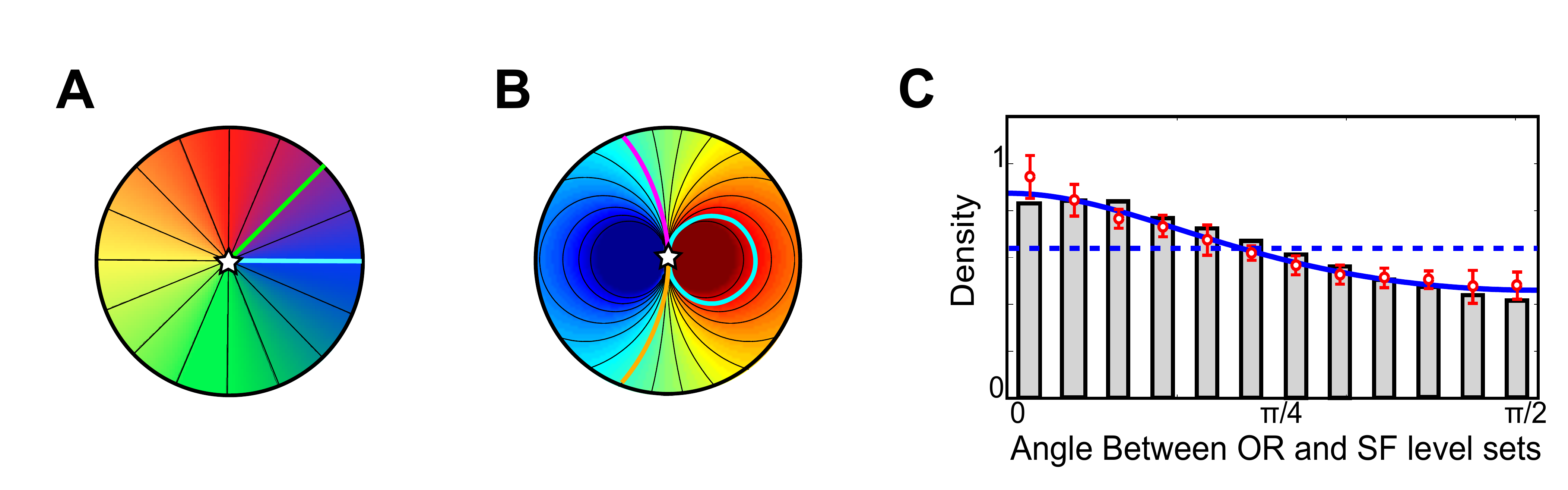}
	\caption{The pinwheel (A) and dipolar (B) architectures. Two arbitrary level sets are represented: in the pinwheel topology, they connect the singularity (star) to the boundary, in the dipolar topology, they are either a single arc connecting the singularity to itself or made of two arcs connecting the singularity to the boundary. (C): Probability density of the angle between iso-OR and iso-SF lines. Red: experimental values of the mean (circles) and standard deviation (error bars) across the different cats ($n=4$), within 150 $\mu m$ from the PCs. Blue dashed line: uniform distribution for the dipole with SF map $\gamma$, gray bars corresponds to the angle distribution of $\gamma$ thresholded at $\pm 1.4$. Blue solid line: best fit with the analytically solvable model $\gamma_{\alpha}$ (here, $\alpha = 0.73$).}
	\label{fig:topology}
\end{figure}

The \emph{topological redundancy} of a map is mathematically defined as the maximal number of connected components of the level sets. For instance, the angle-valued map represented in Fig. \ref{fig:topology}A, said to have the \emph{pinwheel topology}, clearly has topological redundancy one since level sets are single arcs connecting a singular point to the boundary of $\Omega$. In contrast, the real-valued map plotted in Fig. \ref{fig:topology}B, said to have a \emph{dipolar topology}, has topological redundancy two since some level sets are made of two disconnected arcs connecting the boundary to the singularity. These two maps are particularly important: indeed, we shall demonstrate that these are the unique topologies that are surjective (i.e., exhaustive) and minimize the topological redundancy. We note that such maps necessarily show singularities, and call \emph{smooth simple maps} those that are continuous except at isolated points. We demonstrate in Appendix~\ref{append:math}:

\begin{theorem}\label{thm:topology}
	Smooth simple maps that are exhaustive and parsimonious enjoy the following universality:
	\begin{enumerate}
		\item Simple smooth maps  $f:\Omega \mapsto \S^1$ that are exhaustive and with redundancy 1 have the topology of the pinwheel.
		\item Simple smooth maps  $g:\Omega \mapsto U$ that are exhaustive and and optimally parsimonious (redundancy 2) at arbitrarily small scales have the topology of the dipole.
	\end{enumerate}
	Consequently, pairs of smooth simple maps $(f,g): \Omega \mapsto \S^1\times U$ satisfying both minimality and exhaustivity at arbitrarily small scales are the PD topology with co-localized singularities. 
\end{theorem}

This theoretical result is very general: it shows a universal property of maps satisfying local exhaustivity and parsimony principles. In particular, in view of our biological problem, shall the OR and SF maps satisfy these two principles, one will necessarily find PD structures in the vicinity of the PCs of the OR map, and this should extend to the representation of other attributes. 

This mathematical prediction has several implications that account for some experimental facts~\cite{ribot2013organization} inconsistent with the orthogonal architectures, including (i) the sharp transition of the SF map at PC locations, and (ii) the non-orthogonal distribution of angles between iso-OR and -SF lines.  

Indeed, PD structures show a globally uniform distribution, with generically a small bias towards alignment for saturating models. In order to show the latter property,  we shall study a simple model of PD architecture, with an SF map chosen in analogy with the electric dipole potential in 2 dimensions. In detail, the PD model is given by the dimensionless maps\footnote{Here, $\Omega$ is assumed to be the unit disc $\{\vert z\vert<1\}$ of the complex plane and $z=re^{i\phi}$ is the complex coordinate. } $\varphi: z \mapsto \arg (z)/2$ and $\gamma: z \mapsto {\rm Re} (1/z)$ or, in polar coordinates for $z=r\,e^{i\phi}$, $ \varphi:z \mapsto \frac{\phi}{2}$ and  $\gamma: z \mapsto \frac{\cos(\phi)}{r}$, respectively for the OR and the SF (see Appendix~\ref{sec:appendix angles} for more details). 
For this specific pair of maps, it is easy to show by direct calculation that the angle distribution is uniform. Although qualitatively consistent with the overall flat distribution reported in~\cite{ribot2013organization}, it does not account for the slight over- (under)-expression of parallel (orthogonal) lines. This is due to the unrealistic sharp divergence of the SF representation of the electric dipole. Biological dipoles shall saturate to a maximal and minimal value at the singularity, and this saturation recovers this bias, as we show in the Appendix \ref{sec:appendix angles}. For instance, a simple generalization of the $\gamma$-map that allows for analytical developments, $\gamma_{\alpha}: (re^{i\phi} \mapsto \frac{\cos(\phi)}{r^{\alpha}})$ with $\alpha<1$, which is less sharp than the $\gamma$ map (but still diverging), reproduces the distribution of angles very accurately as we show in Fig.~\ref{fig:topology}C. The thresholded $\gamma$ map also has the generic property of fitting accurately the distribution (Fig.~\ref{fig:topology}C), as generically do maps with SF saturating at the singularity (see Appendix~\ref{sec:appendix angles}).

These properties of PD architectures tend to point towards the fact that dipoles are consistent with previously reported facts on the behavior of the SF map at PCs. Direct evidences of PD architectures are obtained in a companion paper~\cite{ribot14} using new high resolution optical imaging data to resolve the fine structure of the SF map on cat's V1 area in the vicinity of PCs (see {\it Material and Methods} and Appendix \ref{Methods}). In that paper, we prove that at a majority of PC locations, the SF map is in fact organized as a dipole, by showing (i) that the map displays both a global maximum and a global minimum of the SF representation and (ii) that the thresholded $\gamma$ model (with possible deformations, see Appendix~\ref{append:Coding}) fits with high accuracy the structure of the map in a circular region of $150\,\mu m$ around PCs.

\subsection{Balanced detection of multiple attributes}\label{sec:Balanced}
\begin{figure}[h]
	\centering
		\includegraphics[width=.8\textwidth]{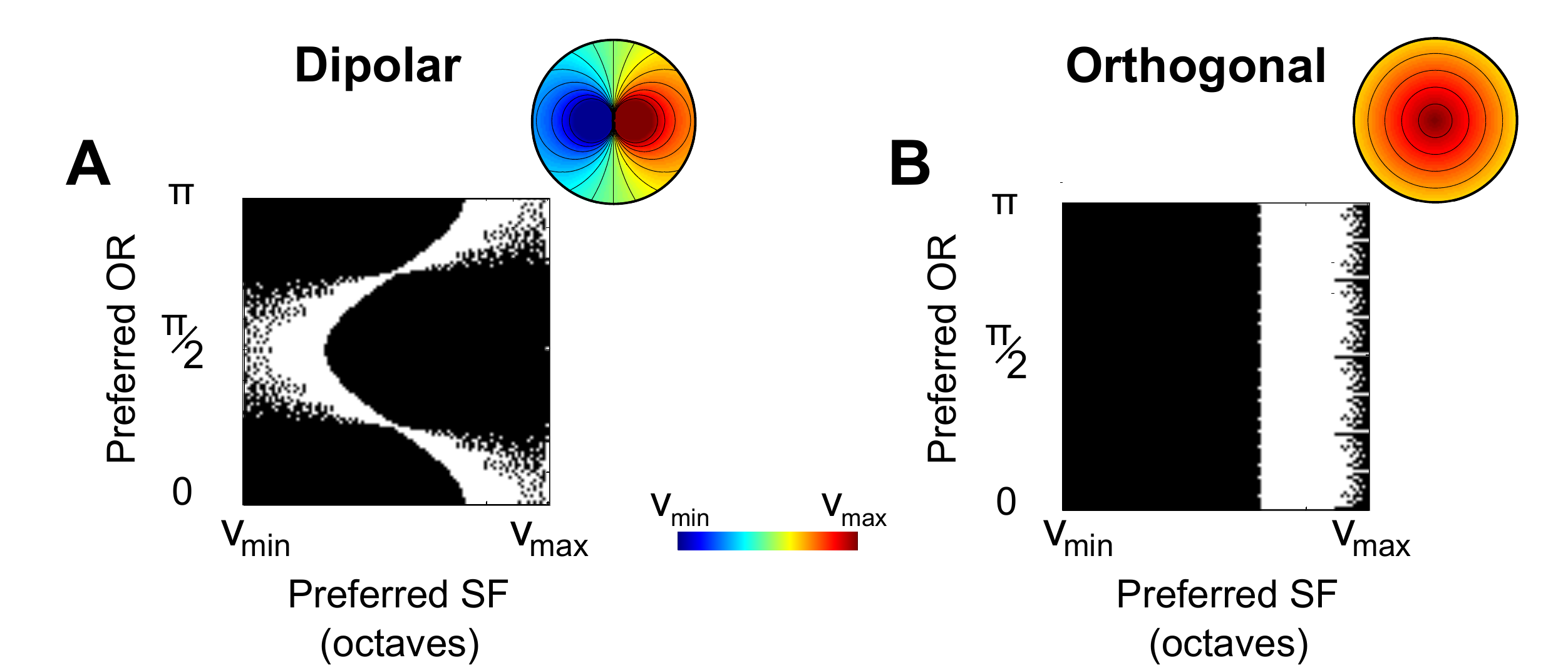}
	\caption{Parameter space coverage near PCs for dipolar and orthogonal architectures. (A): White pixels correspond to preferred couples $ (\theta^*, \nu^*)$ for the OR map $\varphi$ and the SF map $\gamma$ thresholded at the extreme values $\nu_{min}$ and $\nu_{max}$, shown in the top-right corner. (B): The same as (A) for the putative orthogonal architecture with the SF map, corresponding to a local maximum, shown in the top-right corner.}
	\label{fig:pixeldistrMain}
\end{figure}

From the functional viewpoint, the fact that the PD architecture is highly non-orthogonal implies that the sampling of the attributes is not uniform near PCs. One may therefore expect the coding properties of the PD architectures to be very different than in a uniform coverage architecture. Under the uniform coverage assumption, orthogonality of level sets implies that the SF representation reaches a maximum (or a minimum) at the PC and smoothly decays away from the PC. Therefore, in a neighborhood of the PC, only a small portion of the range of SF is represented, contrasting with the PD architecture which is exhaustive. In particular, full representation of both high and low SFs in the orthogonal architecture would necessitate at least two PCs, one corresponding to a maximum of SF representation and the other one a minimum. At one given PC with orthogonal topology (say, corresponding to a maximum of the SF representation), it is likely that stimuli with high SF will be well encoded regardless of their orientation, but may be blind to low SF stimuli. In contrast, the PD structure represents both high and low SFs, but different SFs are associated to distinct ranges of OR (see Fig. \ref{fig:pixeldistrMain}). Overall, in both architectures, we find cells with preferred OR and SF covering the same proportion of possible stimuli, but with a different structure in the parameter space. Across all possible stimuli, it is therefore a priori unclear whether one of the architectures is more efficient in encoding an information. In order to investigate this question, we numerically computed the normalized errors made on encoding the OR ($\epsilon_{\theta}$) and the SF ($\epsilon_{\nu}$), using a simple procedure of coding and decoding the information, in a model of PD and orthogonal architecture fitted to the optical imaging data (see {\it Materials and Methods} and Appendix~\ref{append:Coding}). 

\subsubsection{A perceptual trade-off}

In the coding procedure, the only parameters that are not evaluated from the fits to optical imaging data are the shape of the tuning curves that, in the classical models (wrapped Gaussian for OR or Gaussian for SF) depend on a single parameter, the tuning width, or Full-Width at Half-Height (FWHH). We denote by $y$ the OR tuning width and by $w$ the SF tuning width. While the OR selectivity $y$ has been well estimated in the vicinity of pinwheels, ($y_{exp}=80\degree$ for OR near the PCs, see e.g. \cite{rao1997opticaily}), this is not the case of that of the SF map. In \cite{ribot2013organization}, a global evaluation on the whole map of the SF tuning width was reported (median $\pm$ mad $ w_{exp}^{all} = 2.48 \pm 0.19$ octaves). 

Numerical results using these tuning widths show that the accuracy of OR detection in the dipolar structure is significantly degraded compared to the orthogonal one $\epsilon_{\theta}= 3.4 \pm 1.2 \%$ vs $0.06 \pm 0.01 \%$  (median normalised error $\pm$ mad; Mann-Whitney-Wilcoxon test, $p < 10^{-3}$). Nevertheless, the SF detection is vastly improved, $\epsilon_{\nu}=8.1 \pm 1.8 \%$ vs. $20 \pm 3 \%$ ($p < 10^{-3}$). The conjoint accuracy of SF and OR detection is therefore globally improved in the dipolar structure, $\epsilon_{tot} = 14 \pm 2 \%$ vs. $20 \pm 3 \%$ ($p < 10^{-3}$). This first result tends to show that local dipolar architecture is not only comparable, but can even outperform the orthogonal architecture.

\begin{figure}[h]
	\centering
		\includegraphics[width=1.\textwidth]{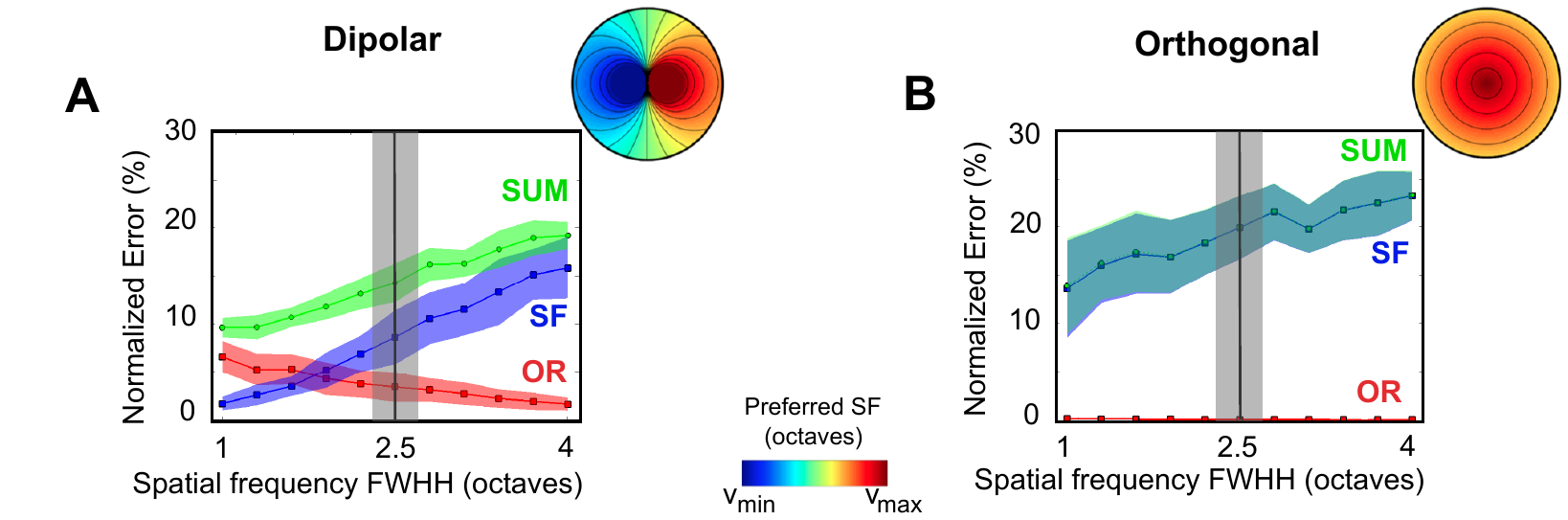}
	\caption{OR and SF tuning widths optimizes balanced detection at PCs. (A): Median error (in $\% \pm$ mad) in the detection of OR (red), SF (blue) and OR+SF (green), for the dipolar structure as a function of the FWHH for the SF tuning curve (in octaves). The FWHH for the OR tuning curve is set at $y_{exp}=80\degree$. The gray bar corresponds to the experimental variability of the SF tuning \cite{ribot2013organization} of the whole map. One example of the SF maps considered is shown in the top-right corner. (B): The same as (A) for the orthogonal architecture.}
	\label{fig:tradeoff}
\end{figure}

These performances in evaluating the correct stimuli are expected to vary depending on the SF tuning width. In order to investigate the importance of this dependence on selectivity, we therefore simulated the model for distinct values of the SF tuning width (Fig. \ref{fig:tradeoff}), both for the PD and for the orthogonal architecture. Depending on the architecture, we observed very distinct qualitative behaviors of the errors when the SF tuning width is varied. As expected, the error in perceived SF (blue curve) increases with the SF tuning width in both architectures. In the orthogonal architecture (Fig. \ref{fig:tradeoff}B), the error in perceived SF remains much higher than the one in perceived OR (Mann-Whitney-Wilcoxon test, $p< 10^{-3}$), and the improved joint accuracy of the detection of the dipolar architecture over the orthogonal one mentioned for one typical values of the SF tuning width above, persists in the whole range of $w$ considered.

In the dipolar architecture, a striking phenomenon arises: in contrast to the SF error, the error in perceived OR (red curve) in the PD architecture decreases, and the SF and OR error curves intersect. Therefore a specific tuning width ensures balanced error in OR and SF perception. However, the specific value of the intersection (mean $\pm$ sd: $\overline{w}=1.73 \pm 0.31$ octaves) is significantly below $ w_{exp}^{all}$ ($Z$-score $= 1.79$, one-tailed $p$-value $p= 0.04$).

\begin{figure}
	\centering
		\includegraphics[width=.7\textwidth]{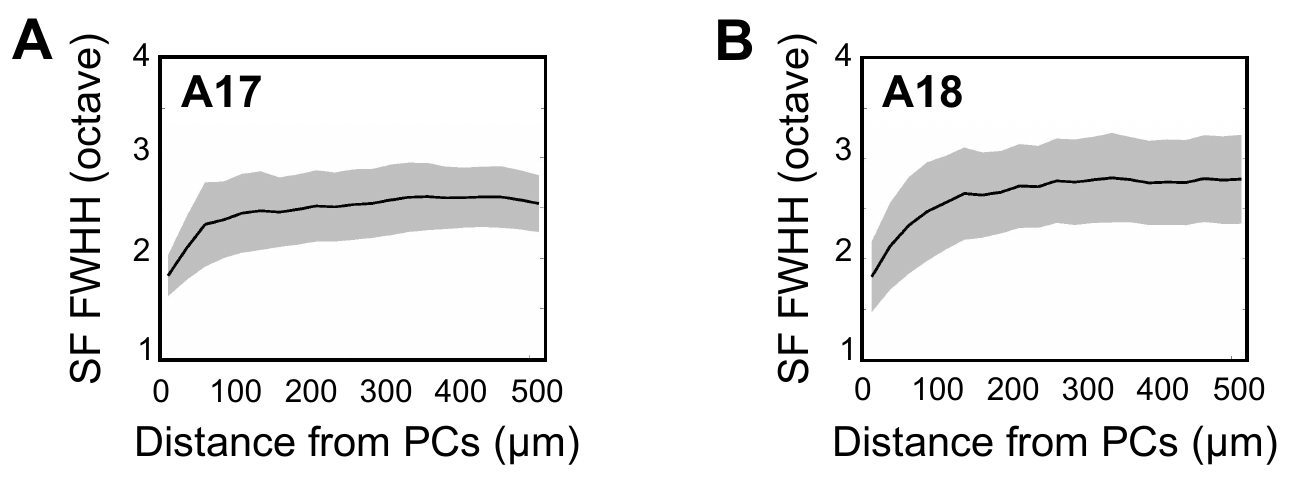}
	\caption{SF tuning width drops close from PCs. (A): Median ${\rm FWHH_{SF}}$ (black curve) $\pm$ mad (gray) as a function of the distance from PCs in A17. Each point corresponds to a distance of 25 $\mu$m. (B): The same as (A) for A18.}
	\label{fig:prediction}
\end{figure}

Observing that the PD maps we considered in our study are valid only locally around the common singularity, a balanced detection hypothesis in OR and SF would thus suggest that, close to PCs, the SF tuning curve shall be sharper. We therefore came back to our experimental data and investigated finely the SF tuning width and its dependence as a function of the distance to the set of PCs.  Strikingly, our data showed that SF tuning width sharply drops close to PCs (at a distance of around $100~ \mu m$, see Fig. \ref{fig:prediction}), reaching the value (median $\pm$ mad)  $ w_{exp}^{PC17} = 1.83 \pm 0.20$ octaves in A17 and  $ w_{exp}^{PC18}  = 1.83 \pm 0.24$ octaves in A18, within $25 ~\mu m$ from PCs. These measurements are therefore consistent with the balanced detection determined theoretically  (A17: $Z$-score $= 0.23$ two-tailed $p = 0.82$; A18: $Z$-score $= 0.21$, $p = 0.83$).

Notice that, on the other hand, the joint error in OR and SF detection (green curve) linearly increases with increasing $w$ and is thus minimal for low $w$ ($1.0$ octave here). Therefore this argues that selectivity properties near PCs, rather than improving the joint perception, favor a trade-off in the detection precision of the two attributes. 

\subsubsection{Robustness of the results for the trade-off}

In this analysis, we used a model of dipole which, while fitted with accuracy to the optical imaging data, imposes a specific radial profile of the SF map. One may therefore wonder first of all whether the emergence of a balanced detection between OR and SF qualitatively 
\begin{figure}[h]
	\centering
		\includegraphics[width=.6\textwidth]{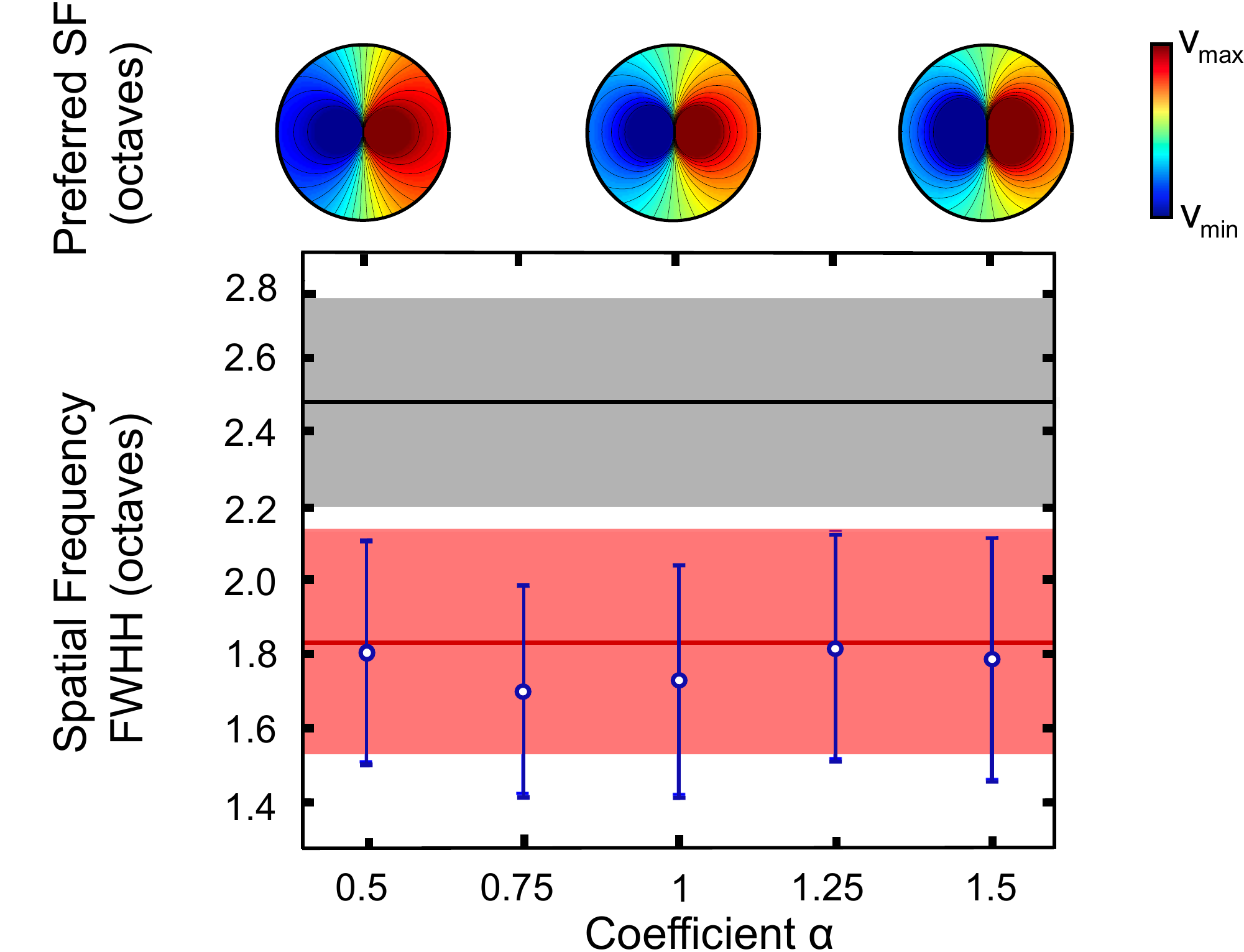}
	\caption{Balanced detection over the class of dipolar architectures $\gamma_{\alpha}$. (Bottom) Blue: Mean (circles) and standard deviation (error bars) for the value of the FWHH of the SF tuning curve corresponding to the tradeoff (intersection between $\epsilon_{\theta}$ and $\epsilon_{\nu}$ curves) for different values of the parameter $\alpha$. Black line and grey horizontal bar: experimental variability (median $\pm$ standard deviation) of the SF tuning width when the whole V1 is taken into account. Red: the same experimental results when the analysis is restricted within 25 $\mu m$ from PCs in A17 (Fig. \ref{fig:prediction}). (Top): Representation of the SF map in the pinwheel area, respectively for $\alpha = 0.5, 1, 1.5$.}
	\label{fig:alphatrend}
\end{figure}
and quantitatively persists across different dipolar architecture. Moreover, it is interesting to investigate how much the related numerical prediction concerning the SF tuning width is robust under deformations of the models. One natural generalization of the model, introduced above, consists in considering nonlinear changes in the SF map scale compared to the physical distance. Therefore we considered dipoles of the form $\gamma_{\alpha}$, with $\alpha \neq 1$, which keep the same topology but affects the shape of the dipole. As already shown in Fig. \ref{fig:topology}C, these deformations can improve ($\alpha < 1$) or degrade ($\alpha > 1$) fits of the angle distributions. It would be possible that these can also affect the possibility of a balanced detection, or at least, if the trade-off persists, modify the precise value of the tuning width corresponding to the curves intersection. 

To this end, we allows for angular and asymmetry deformations of this model for different values of $\alpha$ and fitted the parameters to the data. That way, we obtained different best-fitted dipoles that visually look relatively different (see Fig. \ref{fig:alphatrend}, top panel). These different best fitted architectures were used to test the error made on the OR and SF evaluation. In all cases, curves of errors in OR and SF keep the same monotonicity as the SF tuning width was varied, and intersect. The values of the intersection were computed and reported in Fig. \ref{fig:alphatrend}, bottom panel. Independently on the parameter $\alpha$, they are not consistent with $w_{exp}^{all}$ ($Z$-score $> 1.60$, one-tailed $p$-value $< 0.06$), but actually always compatible with $w_{exp}^{PC17}$ ($Z$-score $< 0.32$, two-tailed $p$-value $> 0.75$). 

Of course, the fit of the whole class of models $\gamma_{\alpha}$ with the same experimental data, coherently plays an important role in constraining the predicted value for $\overline{w}$. These results therefore tend to stress that not only the presence of a trade-off between the two normalized errors is a natural feature of the PD architecture, but also that, regardless of the precise profile of the theoretical map, this fact takes place for SF selectivities in perfect agreement with the biological measurement of SF selectivity near PCs.

\section{Discussion}
This study showed that optimizing simple criteria strongly constrains the layout of maps, and that these layouts can provide specific coding capabilities. We concentrated here on two principles, local exhaustivity and parsimony, which imply that both maps shall display co-localized singularities, around which the orientation map is organized as a pinwheel and the spatial frequency map as a dipole. While pinwheels were identified since decades~\cite{bonhoeffer1991iso}, dipoles were never observed in previous studies. In a companion paper~\cite{ribot14}, high resolution optical imaging made it possible to observe these topologies and validate quantitatively the presence of PD singularities. It is striking that both maps satisfy the same optimality properties near the same singularities. 
At the scale of the whole map, fine detection of local characters in the visual scene makes optimal parsimony not desirable: it is rather important to be able to have accurate detections at several places of the visual scene. Principles at play in the architecture of the whole map shall therefore take into account this necessity, as well as some invariance principles~\cite{wolf1998spontaneous,kaschube2010universality} that predict the density of singularities with respect to a map typical scale. Moreover, purely global criteria such as continuity and coverage are often not sufficient to reproduce quantitatively V1 architectures~\cite{reichl2012coordinated-1,reichl2012coordinated-2,keil2011coverage}. It is likely that the overall structure of V1 emerges from a combination of local criteria, invariance principles and global continuity and coverage optimizations.

Locally around the singularity, we showed that the maps organization has important implications in coding capabilities. One outstanding result was the observation that our model allows for a balanced detection of the OR and the SF, but for tuning widths smaller than the value reported in the literature. This fact did not depend tightly on the choice of the model: it was consistently obtained qualitatively and quantitatively over a class of functions fitted to the experimental data. Strikingly, while estimating the tuning width in the vicinity of pinwheels, we observed a sharp decrease perfectly consistent with the hypothesis of a trade-off between the detection accuracy of OR and SF. This is a surprising phenomenon: the sharpening of the SF tuning curve contrasts with the well-documented broadening of orientation selectivity near pinwheels~\cite{ohki2006highly, rao1997opticaily}. It is interesting to note that sharpening cannot be an artifact of the presence of a singularity, at which subsampling related to the imaging resolution may induce rather a broadening of the selectivity as in the OR case. This significant sharpening effect of the SF selectivity at the singularity points towards the fact that neurons at PC locations develop specific properties, in the vein of their singular resistance to monocular deprivation~\cite{crair1997relationship} or sensitivity to OR adaptation~\cite{dragoi2001foci}.

From a biological viewpoint, it would be worthwhile to investigate whether this study can be extended beyond the case of the OR and SF maps of the cat. This would necessitate to record functional maps for other attributes, and check whether these satisfy exhaustivity and parsimony of representation, if the singularities of OR-SF maps are special points of these maps, and if selectivity properties ensure balanced detection. It should not be surprising that other principles constrain the layout of other maps. In particular, the direction preference map is not exhaustive near pinwheels, and is discontinuous~\cite{shmuel1996functional}, because it is constrained to the orientation map. Another example is given by the ocular dominance map that was shown to be orthogonal to the orientation~\cite{bonhoeffer1995optical} and to the binocular disparity map~\cite{kara2009micro}, which may be optimized to ensure the even representation of visual input from both eyes. 

From a mathematical modeling viewpoint, this study opens exciting perspectives. Indeed, the good agreement between the theoretically derived model and new data constitute an encouraging step towards the development of more complex models that could account for higher order visual areas that respond to more complex features.

\section{Material and Methods}

\subsection{Optical Imaging data} 

High-resolution optical imaging in the cat visual cortex areas A17 and A18 was used to extract OR and SF preference maps. All experiments were performed in accordance with the relevant institutional and national guidelines and regulations. Experiments were conducted on 4 anesthetized young adult cats, and intrinsic signals were recorded with a resolution of 5.9 to 15.3 $\mu$m/pixel. Drifting-grating rotating stimulation were used~\cite{kalatsky2003new} with thirty SFs presented in random order. For each pixel, the modulation of the signal induced by the rotation of the gratings was interpolated via a least-square method with a cosine function whose phase was equal to the preferred OR at this pixel and whose frequency was equal to half the frequency of rotation. Then, at each pixel, the intrinsic signals were interpolated with a difference of Gaussians function and three parameters were extracted: the preferred SF, the tuning width and the error-of-fit. In depth description of the experimental protocol is detailed in the Appendix~\ref{Methods}.

\subsection{Models of PD architectures and fits}
In order to investigate the different efficiencies of the PD and orthogonal architectures, we have simulated circular regions $\Omega$ of radius $R = 50$ pixels. Each pixel of these discs represents a set of neurons responding to a specific range of OR and SF, distributed around the singularity at the origin. The amplitude of their response to varying external stimuli produces graded responses, peaked at a specific value (the quantity represented in the corresponding functional map) and well approximated by the product of a wrapped Gaussian function, for the OR coordinate, and a Gaussian function, for the SF one (the tuning curves, see Fig. \ref{fig:coding}B)~\cite{baker2005cortical}. 

The orientation map has been defined as half of the polar angle, modulo an arbitrary phase. For the dipolar SF map we studied the class of functions $\gamma_{\alpha}$, saturating at extreme values and incorporating possible angular and shape deformations. For the orthogonal case, we considered a rotational invariant map, linearly decreasing from a maximum value located at the origin. 

The realistic range of values for the free parameters of the PD model have been evaluated by fitting to the optical imaging data (restricting to fits with coefficient of determination $> 0.8$, using Matlab\textregistered ~function {\it regress}). The same data have been used to estimate the slope of the radial SF decay parameter in the orthogonal model. For both PD and orthogonal cases, we have simulated 50 different couples of OR and SF maps, differing each other for the parameter sets chosen in these intervals. Details of the models and functions used are provided in Appendix~\ref{append:Coding}.

\subsection{Coding and Decoding an information}

\begin{figure}[h]
	\centering
		\includegraphics[width=0.75\textwidth]{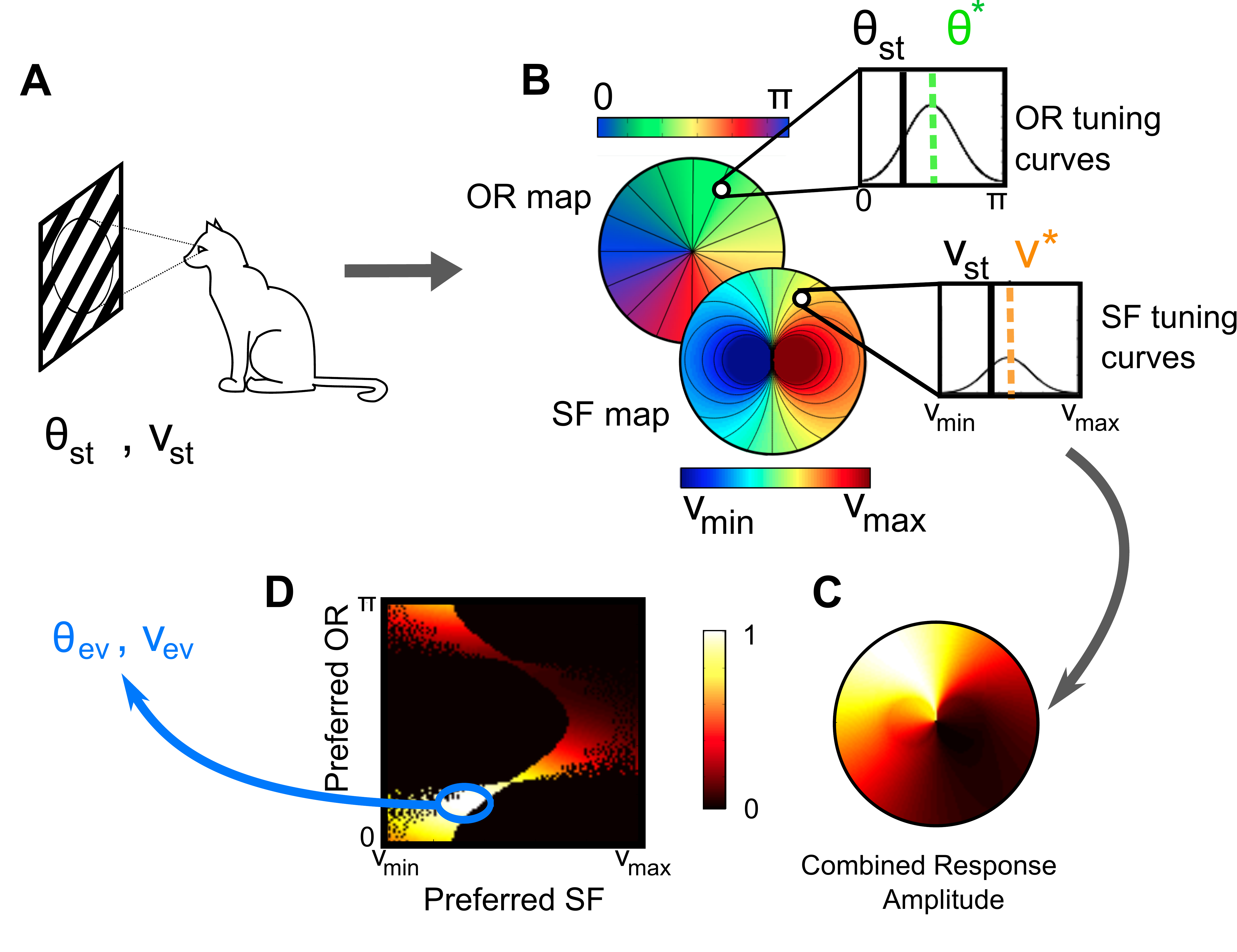}
	\caption{Scheme of the visual perception strategy studied in the paper. (A): The visual external stimulus is represented by a grating defined by the OR and SF values $(\theta_{st},\nu_{st})$. (B): For a given couple of OR and SF local architectures in a disc centered at the common singularity, each cortical location responds to the external stimulus according to its associated OR and SF tuning curves with preferred values $(\theta^*,\nu^*)$. The minimal and maximal values of preferred SF for a given map are noted as $\nu_{min}$ and $\nu_{max}$. (C): The collection of all the combined OR-SF normalised responses of the selective neurons is represented into the cortical area around the pinwheel. (D): The same set of responses is represented in the parameter space (OR, SF). The final estimation $(\theta_{ev},\nu_{ev})$ of the external stimulus, associated  to the local area around the pinwheel, is derived through a weighted average within the $10 \%$ highest response amplitudes.}
	\label{fig:coding}
\end{figure}

In order to evaluate coding capabilities of PD and orthogonal architectures, we used the coding and decoding algorithm represented in Fig. \ref{fig:coding}. A grating stimulus with OR $\theta_{st}$ and SF $\nu_{st}$ (Fig. \ref{fig:coding}A) produces a map of activity over $\Omega$ (Fig. \ref{fig:coding}C). These activity maps are very similar to those obtained in our optical imaging experiments \cite{ribot2013organization, ribot14}. 
The actual algorithm used by the brain to extract from this pattern of activity an estimation of the stimulus attributes is largely unknown. Here, we consider a simple estimation procedure: the attributes of the stimulus are evaluated as the center of mass of the responses in the (OR,SF) plane (Fig. \ref{fig:coding}D). This provides an estimated value $(\theta_{ev},\nu_{ev})$ with relative errors denoted $(\epsilon_{\theta},\epsilon_{\nu})$. For each fixed architecture, a distribution of errors  $\epsilon_{\theta}$ and $\epsilon_{\nu}$ is obtained by varying $(\theta_{st},\nu_{st})$. The errors $\epsilon_{\theta}$ and $\epsilon_{\nu}$ are obtained by normalizing with the respective total interval length; a total error is defined as $\epsilon_{tot} = \epsilon_{\theta}+\epsilon_{\nu}$. Simulations were done considering 100 random external stimuli, for each PD and orthogonal architecture. Appendix \ref{append:Coding} describes in more detail the algorithms and statistical analyses.

 \begin{acknowledgments}
The authors thank Timoth\'ee Coulais for animal care and Chantal Milleret and Sidney Wiener for discussions and comments on the manuscript. The research was partly supported by the Fondation Louis D and the Luz Optique Group, the European Community (Marie Curie International Reintegrating Grant), the Foundation Berthe Foussier (to JR) and the CNRS PEPS PTI Program (NeuroGauge Project to AR and JT). The funders had no role in study design, data collection and analysis, decision to publish, or preparation of the manuscript.
 \end{acknowledgments}

\appendix
\section{Pinwheel-Dipole architecture minimizes geometric redundancy}\label{append:math}
In this section, we analyze mathematically the topology of local representations of OR and SF and investigate their properties. We will be particularly interested in the parsimony and exhaustivity of the representations of these maps. We will show that the pinwheel architecture optimizes the geometric redundancy of representation of a periodic quantity (the orientation), while the dipole optimizes the geometric redundancy of representation of a non-cyclic variable (the spatial frequency). These results go beyond the example of OR and SF representations: a number of sensory modalities have a local character, e.g. touch or vision. Some of the neurons processing this local information have receptive fields that reflect the local domains of sensor cells. In general these receptive fields shall have access to other characteristics of the stimuli, and fine perception necessitates that brain areas reproduce the same characteristics many times, in several places: in our context, early visual cortex of the cat shows quasi-periodicity and continuity of the representation. We will establish that the minimization of redundancy in local domains (i.e. fundamental cells of the quasi-periodic maps) explains most of the experimentally observed structure.

In detail, we show in the case of the OR representation (corresponding to a continuous map with value in a circle) that minimal redundancy 1 is equivalent to the pinwheel architecture. As for SF, we prove that, to represent a real variable which varies in an open interval (or equivalently, from $-\infty$ to $+\infty$), several constraints arise: i) the map must have singularities (points where the value is undefined), and ii) near such singularities of the map, the dipolar architecture is the unique structure that is compatible with continuity, and has minimal redundancy.

It is remarkable that the pinwheel-dipole architecture observed in the cat's visual cortical areas 17 and 18 in the present work is the unique scale-invariant topology achieving minimal representation of a pair $(\theta, \nu)$ for the the circular variable $\theta$ and the variable $\nu$ taking values in an open interval.

\subsection{Topological Redundancy, Pinwheel and Dipole Topologies}

Since the OR and SF maps are quasi-periodic, we limit our study to a fundamental domain $\Omega$, having the topology of a disc. Without loss of generality, we will hence consider that $\Omega$ is disc centered at the origin $0$, and denote by $\Gamma$ the boundary circle of $\Omega$. Coordinates in this disc are denoted $(x, y)$. The letter $u$ denotes the scalar variable to be represented. In one dimension, there are only two types of smooth domains, the open interval $U$ (also isomorphic to the real line $\R$), and the circle $\S^1$ (a closed periodic interval). A cortical map $f$ is an application from $\Omega$ on $\S^1$ (e.g., the OR map) or $U$ (e.g., the SF map). In $\S^1$ we shall generically choose a coordinate $\theta$ defined modulo $\pi$.

We now formally define pinwheel and dipole topologies, examples of which were represented in Fig.~\ref{fig:topology}.
\begin{definition}
	We consider a map $f:\Omega\mapsto \S^1$ and $g:\Omega\mapsto U$. We say that:
	\begin{enumerate}
		\item the $f$ has the \emph{pinwheel topology} if it has a singularity (say, without loss of generality, located at the origin $0$) and each level set connects the singularity to the boundary $\Gamma$ of $\Omega$. Figure~\ref{fig:topology}A represents a typical example\footnote{Any map with the pinwheel topology is equivalent to that topology in the sense of homotopy, i.e., there exists a continuous and invertible map transforming the map into $\Omega: r e^{i\phi}\mapsto\frac{\phi}{2}$.}.
		\item $g$ has the \emph{dipolar topology} if (i) it has a singularity, (ii) its level sets are either made of single closed arcs connecting the singularity to itself or two arcs connecting the singularity to the boundary, and (iii) it has exactly two disjoint families of loops connecting the singularity to itself. Fig.~\ref{fig:topology} B is a typical example of such topology.
	\end{enumerate}
\end{definition}
 
As noted in the main text, pinwheel topology has geometric redundancy 1, and is moreover scale invariant in the sense that both exhaustivity and redundancy $1$ persist in any smaller neighborhood of the singularity (e.g., in Fig. \ref{fig:topology}A, for any smaller circle around the singularity). Similarly, dipoles have topological redundancy 2, and scale invariance property: all levels are represented arbitrarily close from the singularity and the geometric redundancy is two on any circle, arbitrarily small, around the singularity.

\subsection{Minimal representations of orientations and universality of the pinwheel}

In this section, we show that pinwheel architecture is the unique minimizer of geometric redundancy allowing exhaustive representation of all orientations in $\Omega$. The proof proceeds by showing that the representation cannot be minimal without presenting at least one singularity in $\Omega$, and that all level sets connect that singularity to the boundary.
 
\begin{theorem}\label{pro:pinSing}
	A continuous surjective map $f: \Omega \mapsto \S^1$, with redundancy 1, has the topology of the pinwheel.
\end{theorem}

\emph{Proof:} Let $\theta_0\in f(\Gamma)$ be a fixed value. The level set $f^{-1}(\theta_0)$ is a connected set $C$ intersecting $\Gamma$. The complementary of $C$ in $\Omega$ cannot have strictly more than two components: indeed, if that was the case, the continuity of $f$ would imply that there exist at least two connected components of $f^{-1}(u)$ for a value of $u$ near $\theta_0$ contradicting the redundancy 1 assumption.

If the level set $C$ splits $\Omega$ into two disconnected components: $\Omega_{1}$ in which the values of $f$ are larger than $\theta_0$ in the neighborhood of $C$,  and $\Omega_2$ where $f$ is locally smaller. We denote by $I_1=[\theta_0,\theta_1]$ the interval of values covered on $\Omega_1 \cup f^{-1}(\theta_0)$ (considering a continuous in version of the argument, i.e. we have possibly $\theta_1>\pi$ ) and $I_2=[\theta_2,\theta_0]$ on $\Omega_2\cup f^{-1}(\theta_0)$ (with, again, $\theta_2$ possibly smaller than $0$). Since both sets are compact, these values are reached. Then if $\theta_1\geq \theta_2+\pi$, the redundancy is greater than $2$, and if not there is an interval of $\S^1$ that is not covered by $f$ contradicting surjectivity. 

The level set therefore does not splits $\Omega$ into disconnected components. There are two possibilities: either $C \neq \Gamma$, discarded using the same argument as above, or $C$ enters the interior of $\Omega$ and necessarily ends at one point $z^*$. Let us fix two points $z_1$ and $z_2$ outside of the level set, and two arcs $\gamma_1$ and $\gamma_2$ connecting them, and such that $\gamma_1$ crosses $C$ and $\gamma_2$ does not. The map $f$ restricted to $\gamma_1 \cup \gamma_2$ therefore covers $\S^1$. This is true arbitrarily close to $z^*$, implying that $z^*$ is a singularity and all level sets converge to it.

Proving that the map has the pinwheel topology only amounts showing that all levels are represented along the boundary.
The map $f$ restricted to $\Gamma$ is continuous. If it does not cover $\S^1$ then along $\Gamma$, $f$ takes necessarily twice the same value, contradicting redundancy 1 (since we have seen that any level set starting from the boundary reaches necessarily the singularity). 
Therefore, all levels are represented on the boundary $\Gamma$ and reach the singularity: the map has the topology of the pinwheel.
\qed

The topology of the pinwheel is therefore universal: any map minimizing redundancy of representation has this topology. An analogous property is now demonstrated for the representation of SF. 
 
\subsection{Minimal representations of the spatial frequency and universality of the dipole}
We examine now the topology of exhaustive and parsimonious continuous maps taking value in an open interval $U$.
The logarithm of the spatial frequency is a non-periodic real quantity, and within the cortex, neurons respond to a specific range. This interval is not universal: it may vary across animals, and within the same animal across different regions. In that view, the representation of SFs cannot be considered within a closed interval.
This justifies the choice of $U$ as an open interval, which is mathematically equivalent to considering continuous representations of the whole real line. We will therefore consider in the rest of the section $U = \R$. Continuous maps defined on $\Omega$ with no singularity, cannot be surjective on $\R$. Indeed a continuous map on a compact set reaches finite maximum and minimum values. Therefore we need to consider maps that have discontinuities. Simplest maps are those with one single singularity (e.g., at zero). We shall from now on consider continuous maps of the pointed disc ($\Omega^\times = \Omega \setminus \{ 0 \}$). We aim at characterizing maps satisfying the assumptions: 
\renewcommand{\theenumi}{(H\arabic{enumi})}
\begin{enumerate}
	\item \emph{Regularity:} the map is smooth in the sense that all level sets are smooth curves.
	\item \emph{Exhaustivity:} the map is surjective in any neighborhood of $0$;
	\item \emph{Parsimony:} the topological redundancy is minimal at any scale, i.e. there exist arbitrarily small topological discs around the singular point on which the representation achieves the minimal geometric redundancy.
\end{enumerate}
 
The dipole is continuous (H1) and exhaustive (H2). We now prove that it is also parsimonious (H3).
 
\begin{proposition}\label{pro:Redundancy}
	There is no continuous surjective map $g: \Omega^{\times} \mapsto \R$ with geometric redundancy 1.
\end{proposition} 
 
\emph{Proof :} Let $u_0 \in g(\Gamma)$. If the level set of $u_0$ does not intersect the interior of $\Omega$, then by continuity the values in the interior are either all larger or smaller than $u_0$ contradicting the surjectivity. Therefore, the level set enters the interior of $\Omega$. If $g$ has geometric redundancy one, $g^{-1}(u_0)$ is connected. If the singularity at the origin is not in its adherence, it cuts the disc in two parts, one of which containing the singularity. And by symmetry, we can assume that $g$ is larger than $u_0$ on the part $\Omega'$ that does not contain the singular point (and smaller in the other domain). This set $\Omega'$ is compact, the map $g$ reaches a finite maximum on that set, thus it cannot be surjective. The level set of $u_0$ therefore reaches the origin, and the complementary of the level set $\Omega\setminus g^{-1}(u_0)$ is connected. The values of the map, which is continuous on this set, do not intersect $u_0$, therefore are either larger or smaller than $u_0$, contradicting again the surjectivity.\qed

Therefore, minimal representations of the SF have at least geometric redundancy 2. In particular, the dipolar topology is parsimonious (H3). We now show that this dipolar architecture is actually the unique topology ensuring parsimony at small scales. This proof being  slightly more involved, we state the result here, and provide the proof in section \ref{Proof}.
 
\begin{theorem}\label{thm:UnivDipole}
	Any map satisfying assumptions (H1), (H2) and (H3) at arbitrarily small scales around the singular point have the topology of the dipole.
\end{theorem}
 
We conclude that, as is the case of the pinwheel topology for OR representations, not only the dipolar architecture is minimal in the sense of geometric redundancy, it is also the only one that satisfies this property: the dipolar architecture is universal.

\subsection{Proof of the universality of the dipolar architecture} \label{Proof}
 
\begin{figure}
	\centering
		\includegraphics[width=.35\textwidth]{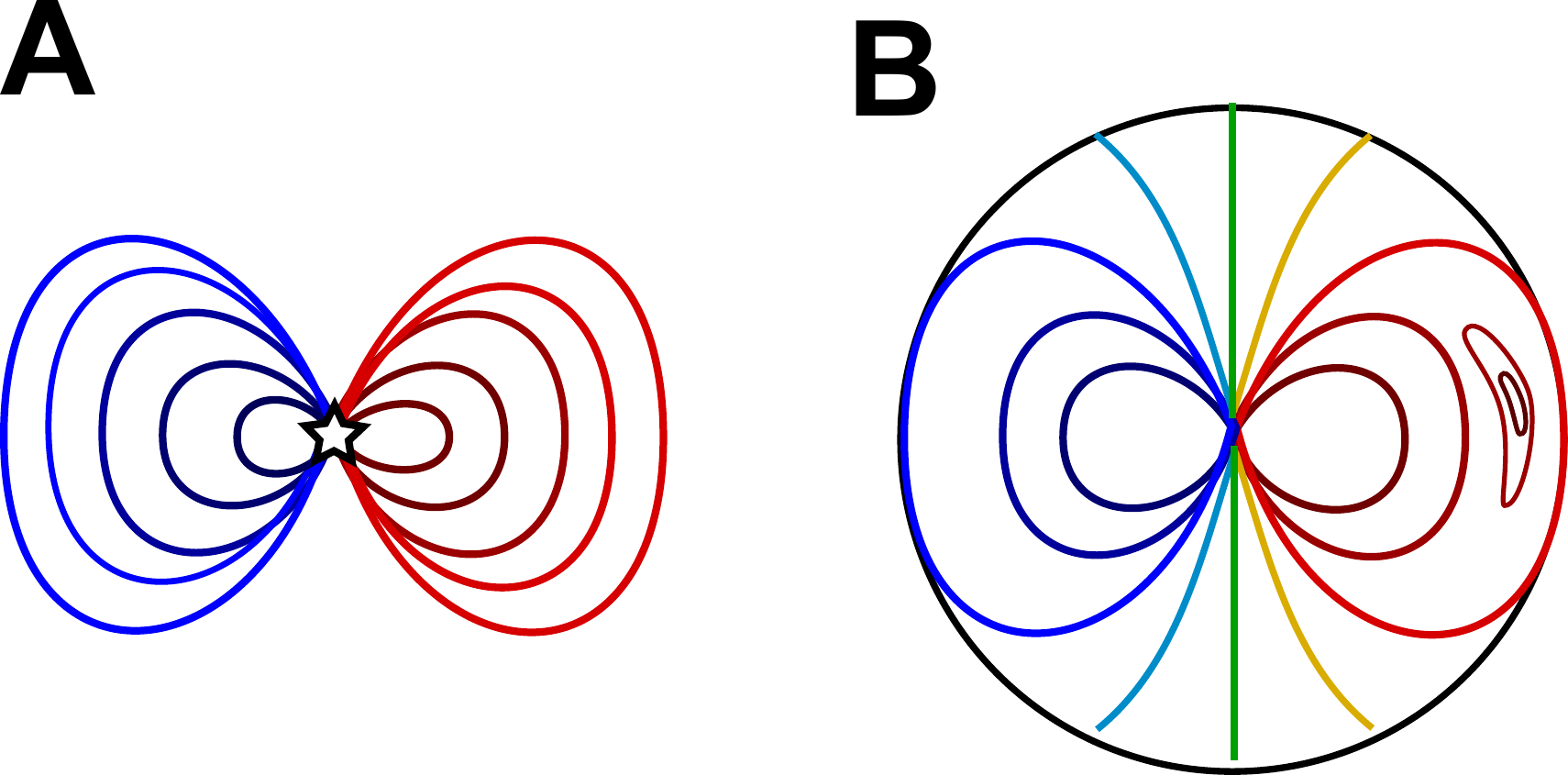}
	\caption{The 8-shapes bouquet (A) and the topology of the dipole (B), with an isolated defect. }
	\label{fig:bouquet}
\end{figure}

We call \emph{8-shapes bouquet} an ensemble of level sets forming two lobes, each of which made of a collection of loops, connecting the singularity to itself.
An 8-shape bouquet is represented in Fig~\ref{fig:bouquet}(A).

The level sets contained inside $\Omega$ are one of four types
\begin{description}
	\item[1)] simple closed curves,
	\item[2)] arcs going from $\Gamma$ to $\Gamma$,
	\item[3)] arcs going from $\Gamma$ to 0,
	\item[4)] arcs with two extremities tending to 0.
\end{description}
 
It is clear that any continuous map $g$ from the pointed disc $\Omega^{\times}$ to the real line $\R$ is continuous on $\Gamma$, and therefore levels connecting to the boundary necessarily represent a bounded set of values. The map hence necessarily displays level sets fully contained inside the disc $\Omega$, i.e. that are either of type 1 or 4. Two types of level sets only add up to the redundancy: these are loops (type 1 level sets) which do not enclose the singularity and type 2 arcs. Indeed, if $C$ is a closed level set that does not enclose the singularity, the level sets of $g$ inside $I(C)$ only cover a bounded interval of values. Within any topological disc containing $0$ and not enclosing $C$, exhaustivity and parsimony shall be satisfied, therefore one will find other connected components of all level sets represented within $I(C)$. In other words, $C$ is an \emph{isolate} in which all levels represented only add up unnecessary redundancy. Similarly, components of type 2 also create isolates (the domain enclosed between the curve of type 3 and $\Gamma$ which does not contain the origin) which only add redundancy. 

Besides this addition to the redundancy, the presence of such isolates may imply the presence of non-smooth level sets corresponding to the local maxima, contradicting the regularity assumption (H1). We therefore consider topologies with no such isolates, and will come back to this assumption at the end of the demonstration. 

Therefore, the only components $C$ of type 1 (i.e. closed curve) considered enclose the singularity in its interior $I(C)$ (which is well defined and homeomorphic to a disc by the Jordan and Schoenflies Theorem~\cite{moise1977geometric}). Within its interior, there cannot be only components of type 1: otherwise, in order to ensure exhaustivity, the map shall show major variations from very large to very low values, which is not possible with redundancy 2. The type 1 component therefore necessarily contains at least one component of type 4.
 
We now analyze the possible presence of these type 4 level sets (closed curves containing the singularity). If $C$ is a component of type 4, by the Jordan-Schoenflies theorem, it encloses a domain $\Delta$ which is homeomorphic to a disc, and contains $0$ in its boundary. It is easy to see that all level sets inside $\Delta$ are necessarily of type 4. 
On one component $\Delta$, $f$ may diverge to $+\infty$ or to $-\infty$, but cannot be doubly unbounded because of the parsimony assumption (shall the map have unbounded positive and negative values, it would present an infinite number of oscillations). These remarks allow now to demonstrate that any possible topology satisfying our assumptions contains a unique 8-shape bouquet.
 
\begin{proposition}\label{pro:Bouquet}
         A map satisfying assumptions (H1), (H2) and (H3) contains a unique 8-shape bouquet, and values of $g$ in each lobes are not overlapping.
\end{proposition}

\emph{Proof:} First of all, as we noted already, surjectivity of the map implies that the map necessarily has components  at least two disconnected components of type 4. Indeed, otherwise all levels not represented on the boundary correspond to curves of type 1. The level sets within the interior of $\Omega$ is then made of nested closed loops circling the singularity, and it is not possible for the map to reach $-\infty$ and $+\infty$ with redundancy two. The map therefore necessarily has components of type 4. Within one component of type 4, the map may reach arbitrarily large (or arbitrarily small) values but not both. Let us assume that it reaches $+\infty$. We denote by $\Delta_+$ the interior of the largest component of type 4 of this family. $\Omega\setminus \Delta_+$ has the topology of an annulus and shall represent all values until $-\infty$, and this is not possible with only arcs of type 1 by continuity. Indeed, let us for instance restrict the map to a topological circle $\Omega'$ containing the singularity and intersecting the boundary of $\Delta_+$. Within $\Omega'$, components of type 1 are not possible because levels cannot cross the connected components of $\Delta$. However, the map within $\Omega'\setminus \Delta_+$ represents all values within a semi-infinite interval $(-\infty, x_{\min}]$. It necessarily has level sets fully included within $\Omega'\setminus \Delta_+$, i.e. components of type 4. 

By the redundancy 2 assumption, the map cannot have more than two lobes made of type 4 arcs. Indeed, at sufficiently small scale around the singularity, an arc of type 4 becomes two arcs of type 3, and by redundancy two assumption, this implies that there is no component reaching the singularity and corresponding to this level. In particular, different lobes cannot represent overlapping values (we would otherwise have 4 connected components for any common value at sufficiently small scale). The same argument may be applied to rule out the presence of strictly more than two lobes. Indeed, if this was the case, consider three levels $a<b<c$ belonging to different lobes. By continuity, there exists an additional connected component for the level set $b$ disconnected from the type 4 arc considered in the lobe, and therefore redundancy 3 at small enough scale, which ends the proof. \qed
 
In order to complete the proof of Theorem \ref{thm:UnivDipole}, we shall now characterize the possible topologies of level sets outside of the 8-shapes bouquet. 
Without loss of generality, we now consider a neighborhood of the singularity intersecting both lobes of the 8-shapes bouquet and in which the map satisfies (H1), (H2) and (H3). We denote by $C_{+}$ and $C_{-}$ the external components of the two branches of the unique 8-shape bouquet converging to $\pm \infty$, $c_{+}$ and $c_{-}$ the respective finite extremal values of $f$ on $C_{+}$ and $C_{-}$. We have shown in the previous proof that necessarily $c_{+}>c_{-}$. 

The rest of the values are now shown to be represented along pairs of arcs of type 3. 

\begin{proposition}\label{lem:8}
	The map $g$ is topologically equivalent\footnote{Mathematically, two maps $f:I(\Omega) \mapsto \R$ and $g:I(\Omega) \mapsto \R$ are said to be topologically equivalent if there exists an homeomorphism (smooth invertible map) $\varphi$ of the open disc $I(\Omega)$ onto itself and a homeomorphism $\psi$ of $\R$ onto itself such that: $f= \psi \circ g \circ \varphi^{-1}$.} to a dipole. In particular, in addition to the 8-shapes bouquet, level sets are arcs of type 3 and the value of $g$ goes monotonically from $c_+$ to $c_-$ on both sides of the bouquet.
\end{proposition}

\emph{Proof:} The bouquet separates the rest of disc in two parts, said left ($L$) and right ($R$). On both parts, the map covers the whole interval $(c_-,c_+)$, and therefore there is exactly one connected component of the level sets of $g$ corresponding to the intervals $(c_-,c_+)$ in $L$ and in $R$. This connected component exists in arbitrarily small neighborhoods of the singularity. Since there is a single curve, it is connects the singularity to the boundary, and the redundancy 2 implies that $f$ varies monotonically on them in $L$ as well as in $R$. We therefore have pairs of arcs of type 3 for every level within $(c_-,c_+)$ and arcs of type 4 in the interval $I_+=(c_+,\infty)$ and $I_-=(-\infty, c_-)$. Let us show that there cannot be additional level sets (apart from small isolated defects). It is clear that additional level sets shall necessarily correspond to values in $I_+$ or $I_-$ that are the only levels which, within $\Omega$, have redundancy 1. By continuity, additional level sets therefore necessarily belong to the interior of the bouquet, and such arcs can therefore only be of type 1 and do not enclose the singularity: these are isolated defects. 
\qed

This concludes the proof of the theorem~\ref{thm:UnivDipole}: apart from isolated defects, the map has the topology of the dipole~\ref{fig:bouquet}(B). An example of the only possible type of defect is added in this picture.

\section{Optical Imaging data} \label{Methods}

High-resolution intrinsic optical imaging was performed in cat visual cortical areas A17 and A18 to record maps of OR and SF. All experiments were performed in accordance with the relevant institutional and national guidelines and regulations (i.e., those of the Collge de France, the CNRS, and the DDPP). Experiments also conformed to the relevant regulatory standards recommended by the European Community Directive and the US National Institutes of Health Guidelines. A complete and in depth description of the experimental protocol is detailed elsewhere \cite{ribot2013organization, ribot14}. 

\emph{Animal Model and Surgical Procedure:} Experiments were conducted on 4 young adult cats aged between 24 and 72 weeks. Animals were anesthetized, paralyzed, and artificially ventilated with a 3:2 mixture of N2O and O2 containing 0.5-1$\%$ isuflurane. Electrocardiogram, temperature, and expired CO2 were monitored throughout the experiment. Animals were installed in the Horsley-Clarke stereotactic frame and prepared for acute recordings. The scalp was incised in the sagittal plane, and a large craniotomy was performed overlying areas 17 and 18 of both hemispheres. The nictitating membranes were retracted and the pupils were dilated. Scleral lenses were placed to protect the cornea and focus the eyes on the tangent screen 28.5 cm distant. 

\emph{Optical imaging}. The cortex was illuminated at 700 nm to record the intrinsic signals. The focal plane was adjusted to 500 $\mu$m below the cortical surface. The optic discs were plotted by tapetal reflection and the center of the screen was moved 8 cm (15$\degree$) below the middle of the two optic discs. Intrinsic optical signals were recorded while the animals were exposed to visual stimuli displayed on a CRT monitor subtending a visual angle of 75$\degree \times 56\degree$. Frames were acquired by CCD video camera (1M60, DALSA) at the rate of 40 frames per second and were stored after binning by $2 \times 2$ pixels spatially and by 12 frames temporally using the LongDaq system (Optical Imaging). Images were acquired with a resolution of 15.3 or 5.9 $\mu$m/pixel.

\emph{{Stimulation}}. Each stimulus consisted of full-screen sine-wave gratings drifting in one direction and rotating in a counter-clockwise manner \cite{kalatsky2003new}. The angular speed of the rotation was 2 rotations per min and the temporal frequency of the drift was 2 Hz. The contrast was set at 50$\%$. Thirty SFs ranging linearly in a logarithmic scale from 0.039 to 3.043 cycle/degree (cpd) were presented in random order. Ten full rotations were presented for each SF. At the end of the last rotation, the first frame of the first rotation for the next SF was presented without interruption. The total duration of the recording was 2.5 hours. 

\emph{Image processing}. Data were pre-processed with the generalized indicator function method \cite{kalatsky2003new} for each SF separately. A low-pass filter with a Gaussian kernel of around 15 $\mu$m half width was also applied for smoothing the data. A Fourier transform was performed on the temporal signal of each pixel for all SFs together. The phase at half the frequency of rotation was calculated to obtain the preferred OR at each pixel \cite{kalatsky2003new}. Then intrinsic signals related to each SF were considered separately. For each pixel, the modulation of the signal induced by the rotation of the gratings was interpolated via a least-square method with a cosine function whose phase was equal to the preferred OR at this pixel and whose frequency was equal to half the frequency of rotation. Magnitude maps for preferred ORs were thus obtained for each stimulus SF \cite{ribot2013organization}. Pixels with negative values, which corresponded to interpolation peaking at orthogonal ORs, were rectified to zero. Then, at each pixel, the intrinsic signals were interpolated with a difference of Gaussians function and three parameters were extracted: the preferred SF, the Òfull-width at half-heightÓ and the error-of-fit.

\section{The angle distribution}\label{sec:appendix angles}

In \cite{ribot2013organization} it has been shown that the experimental distribution of the angle $\psi$ between the level sets of OR and SF maps  near the PC's (i.e., for distances smaller than $150\mu m$) can be roughly approximated with a uniform distribution, with a small significant bias towards small angles. We showed in the main text results that dipoles are good candidates to fit this distribution. We provided an analytically solvable model and simulations with other models, and claimed that essentially, distinct dipolar models can be precisely fitted to the experimental angles distribution. 

Let us start by considering the maps $\varphi: r e^{i \phi} \mapsto \frac{\phi}{2}$ for the orientation and $\gamma_{\alpha}$ for the spatial frequency:
\begin{equation} \label{alphamodel}
\gamma_{\alpha}: r e^{i \phi} \mapsto \cos(\phi)/r^{\alpha}.
\end{equation}
It is straightforward to derive a formula for the intersection angle $\psi$ between the iso-OR and iso-SF lines:
\begin{equation}
\cos \psi= \left| \frac{\sin (\phi )}{\sqrt{\alpha ^2 \cos ^2(\phi )+\sin ^2(\phi )}} \right|. \\
\end{equation}
Noticing that $\phi$ depends only on the angular coordinate $\phi$, the corresponding probability density $P(\psi)$ in a circle around the pinwheel singularity, can be derived from the infinitesimal formula 
\begin{equation}
d\phi = \frac{d\psi}{h'(h^{-1}(\psi))}
\end{equation}
and after normalisation reads
\begin{equation} \label{alphadistr}
	P (\psi) = \frac{2 \alpha }{\pi  \left(\left(\alpha ^2-1\right) \cos ^2(\psi )+1\right)}.
	\end{equation}
The case $\alpha=1$ corresponds to a perfectly uniform distribution for the angles between iso-SF and iso-OR lines. The density $P(\psi)$ is a one-parameter function which can be fitted to the experimental data, and for any $\alpha<1$, the qualitative experimental observation on the shape of the distribution is recovered. Optimizing on $\alpha$ using the best weighted least squares fit, we obtain an excellent fit with the experimental data shown in Fig. \ref{fig:topology}C. The best fitted distribution corresponds to $\alpha=0.73$ (with $95\%$ confidence bounds $(0.70; 0.76)$ and coefficient of determination 0.97).
 
 \begin{figure}[h]
 	\centering
 		\includegraphics[width=.5\textwidth]{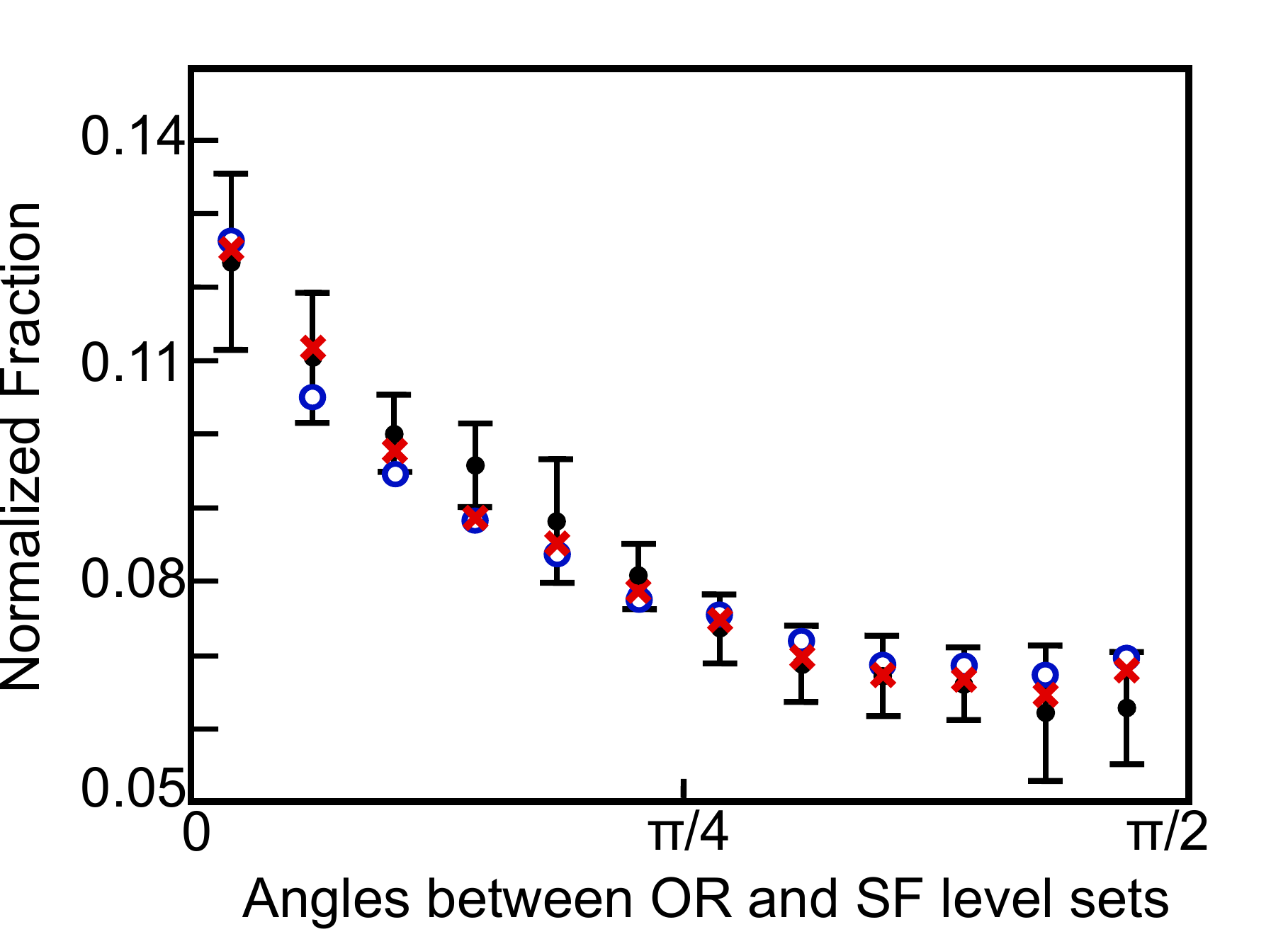}
 	\caption{Distribution of the angles between the level sets of orientation and spatial frequency maps locally around the pinwheels for different models. The black dots and the error bars represent respectively the means and standard deviations with respect to different cats from experimental data. Red crosses and blue circles corresponds respectively to the numerical results for the maps $\gamma_B$ and $\gamma_C$ of Eq. \ref{altmodels}.}
 	\label{fig:fitAnglesTOT}
 \end{figure}

As we mentioned in the text, this is a systematic effect of saturation of the SF map close to the singularity, causing the deformation from a uniform distribution of angles to an excess in correspondence of small $\psi$.

In that sense, it is important to note that this observation is not strongly dependent on the model chosen. The interest of considering the previous $\gamma_{\alpha}$ case, is given by the fact that it allows for analytic developments. Other maps with saturating effects also reproduce qualitatively the experimental observation and allow very good fits with the experimental data. From the biological viewpoint, it is very natural to consider thresholded dipoles: on the one hand, it allows consistency with the data by avoiding a non-biological divergence of the SF map at the singularity, and on the other hand it takes into account the presence of the indeterminacy regions observed in the experimental data in the vicinity of the PCs \cite{ribot14}, which shall not be in our model regions with a high number of level sets in order to avoid biasing the distributions. We consider here, as examples of such maps, the one-parameter models:
\begin{eqnarray} \label{altmodels}
&& \gamma_A:  re^{i\phi} \mapsto \max \left(-A, \min \left(A,  \cos(\phi)\frac{R}{r} \right)\right) \nonumber \\
&& \gamma_B:  re^{i\phi} \mapsto \cos(\phi) \tanh \left(\frac{B\,R}{r}\right) \nonumber \\
&& \gamma_C:  re^{i\phi} \mapsto \cos(\phi) ~{\rm erf} \left(\frac{C\,R}{r}\right)
\end{eqnarray}
where $R$ is the radius of the disc around a PC, and correct dimensions for spatial frequency are thought to be ensured by hidden units in front of each formula. Since in these cases $\psi$ is a function of both coordinates $r$ and $\phi$, it is difficult to find an explicit formula for $P(\psi)$. However the numerical results (with $R=50$ pixels), corresponding to the parameter choice $A = 1.4$, $B= 0.35$ and $C = 0.3$, and represented in Fig. \ref{fig:topology}C ($\gamma_A$) and in Fig. \ref{fig:fitAnglesTOT} ($\gamma_B$ and $\gamma_C$), all show an excellent fit with the experimental data. Therefore, a large family of dipolar architectures can quantitatively reproduce the experimental data. Our experimental techniques do not allow to discriminate between them at the level of the angle distribution; for our purposes, as shown in the next section, the fine structure of the map can be taken into account in an effective way by suitably fixing the parameters of a chosen model.

\section{Neural Coding Model}\label{append:Coding}

In order to test coding capabilities of the SF-OR representation architectures, we have defined a model of OR-SF map based on our theory, and fitted to our experimental observations. We describe in this section the features of the model. 

\subsection{Geometrical descriptions of the OR and SF maps}

We consider a circular region (the pinwheel area, PA) around a PC, and use polar coordinates for the parametrization of the points. For simulations the PA was embedded into a disc of radius $R = 50$ pixels. Two structures have been considered for the SF map: the dipolar architecture observed experimentally in \cite{ribot14}, and a putative orthogonal architecture with a local SF extremum located at the PC \cite{shoham1997spatio,issa2000spatial,issa2008models} consistent with the uniform coverage principle.

\subsubsection{The orientation map:}
Classical neuro-geometrical theory \cite{petitot2008neurogeometrie} describes the OR map in the PA as the half angle of the coordinate, defined modulo $\pi$:
\begin{equation} \label{ORmap}
\theta (r,\phi)=\phi/2+ \phi_0,
\end{equation}
where $\phi_0$ is a phase arbitrarily chosen and different for every pinwheel.
\newline

\subsubsection{The dipolar SF map:}

From the mathematical viewpoint, an idealised map displaying dipolar architecture in the PA is given by the family of functions $\gamma_{\alpha}$ defined in Eq. \ref{alphamodel}, with $\alpha>0$. This $\alpha$-dependent functions exhibits a singularity at the PC (r=0) where it reaches both its minimal and maximal value and shows circular level sets $180 \degree$ apart. 
In order to catch the variability in the angular coordinates, for the numerical studies we allowed moreover deformations of this idealised model, without changing the topology of the level sets. The SF map structure in the PA was then defined as follows:
\begin{equation} \label{SF-def-prel}
\nu_{dip} (r,\phi)= \mathbbm{H}_{\nu_{\min}}^{\nu_{\max}}\left( \mu \left(\frac{R}{r \sqrt{ \cos^2(\phi) \cos^2(\chi) + \sin^2(\phi) \sin^2(\chi)}}\right)^\alpha \cos(\tilde{\phi})  + \nu_0 \right) ,
\end{equation}
where $\mathbbm{H}_{\nu_{\min}}^{\nu_{\max}}$ is the identity map saturating below at $\nu_{\min}$ and above at $\nu_{\max}$, and $\tilde{\phi} = \phi+2 \pi \zeta  \cos (\phi)$,  is an angular deformation allowing architectures in which circular level sets are not $180 \degree$ apart.
The parameters $\alpha, \mu, \chi, \zeta$ and $\nu_0$ allow continuous deformations of the shape of the iso-SF lines. As discussed in the main text and in the previous section, thresholds at a maximum and minimum SF values $\nu_{max}$ and $\nu_{min}$ were added in Eq. \ref{SF-def-prel}. 
In order to evaluate the most reliable values for these parameters to be used for the numerical studies, the model was fitted via a least-square method to the optical imaging data. 
We first focused on the angular variability, neglecting the dependence on $\alpha$ and fixing the simplest case $\alpha=1$.
Only fits with coefficient of determination $> 0.8$ (i.e. fits for which $80 \%$ in the response variable can be explained) were kept, analogously to what has been done in ~\cite{ribot14}. Matlab\textregistered ~function {\it regress} was used to this purpose. 
In a second step, we studied the dependence of our results from the density of level sets in the radial direction, by choosing other values for $\alpha$ and adjusting the other parameters to suitably fit the data. 

The mean value of preferred SF on the global map increases from anterior to posterior region of the cortex ~\cite{tani2012parallel}; however for our purposes, we are mainly interested in the local relative values, and in particular on the difference $\Delta \nu_{extr} = \nu_{max} - \nu_{min}$. The data show that $\Delta \nu_{extr}$ depends on the area of SF indeterminacy cited above. This is consistent with the idealized dipole model that locates extrema at the singularity: the smaller the indeterminacy area, the larger the SF range. The best value after linear regression (mean $\pm$ standard deviation) is given by $\Delta \nu_{extr}= 2.3 \pm 0.8$ octaves ~\cite{ribot14}. Moreover, the data show that (i) $\mu$ can be well approximated as linearly dependent on the difference between the local extreme SF values, and (ii) $\nu_0$ by their average. We consider then $\mu = \mu' \Delta \nu_{extr}/2$ and $\nu_0 = (\nu_{max} + \nu_{min})/2 + \nu'_0$, and all in all we can rewrite 
\begin{eqnarray} \label{SF-def}
&& \nu_{dip} (r,\phi)= \max \left(-\frac{\Delta \nu_{extr}}{2}, \min \left(\frac{\Delta \nu_{extr}}{2}, \tilde{\nu}_{dip} (r,\phi) \right) \right), \nonumber \\
&& \tilde{\nu}_{dip} (r,\phi) = \mu' \frac{\Delta \nu_{extr}}{2} \left(\frac{R}{r \sqrt{ \cos^2(\phi) \cos^2(\chi) + \sin^2(\phi) \sin^2(\chi)}}\right)^\alpha \cos(\phi+2 \pi \zeta  \cos \phi)  + \nu'_0
\end{eqnarray}
The parameters $\alpha, \mu', \chi, \zeta, \nu'_0$ obtained by the fits and successively used in the simulations are listed in Table I.

\begin{table} \label{tab:param}
  \begin{tabular}{| c || c | c | c | c | c |}
  \hline
  $\alpha$ & 0.5 &0.75 & 1 & 1.25 & 1.5 \\
  \hline
  $\mu'$ &  0.7 $\pm$ 0.3 & 0.5 $\pm$ 0.2 & 0.4 $\pm$ 0.2 & 0.4 $\pm$ 0.2 & 0.3 $\pm$ 0.2 \\
  \hline
  $\chi$ &  0.4 $\pm$ 0.5 & 0.6 $\pm$ 0.5 & 0.7 $\pm$ 0.4 & 0.7  $\pm$ 0.3 & 0.7 $\pm$ 0.3 \\
  \hline
  $\zeta$ &  0.0 $\pm$ 0.3 & 0.0 $\pm$ 0.3 & 0.0 $\pm$ 0.3 & 0.0 $\pm$ 0.3 & 0.0 $\pm$ 0.3 \\
  \hline
  $\nu'_0$ &  0.0 $\pm$ 0.5 & 0.0 $\pm$ 0.4 & 0.0 $\pm$ 0.4 & 0.0 $\pm$ 0.4 & 0.0 $\pm$ 0.3 \\
  \hline
  \end{tabular}
  \caption{Parameter of the model fitted on experimental data. The parameter $\nu_0$ is expressed in octaves.}
  \end{table}

\subsubsection{The orthogonal SF map:}

\begin{figure}[h]
	\centering
		\includegraphics[width=.85\textwidth]{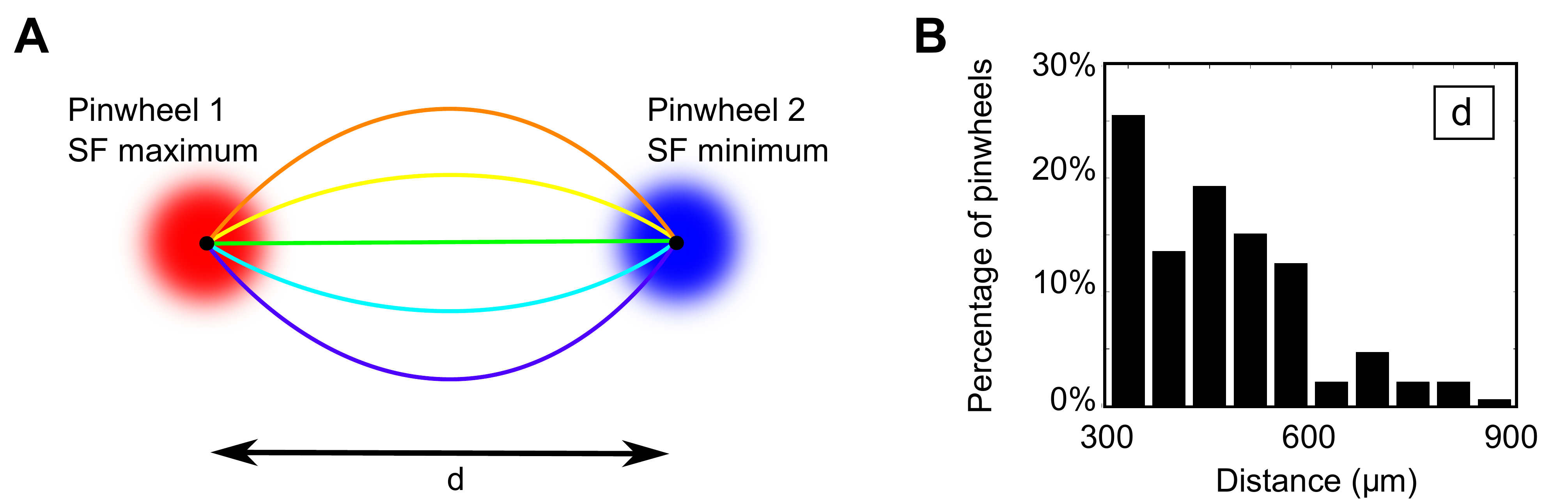}
	\caption{Estimation of the parameters for the orthogonal architecture. (A): Schematic representation of the model used for the orthogonal architecture (adapted from \cite{issa2008models}). In this model, iso-orientation lines (colored) intersect at pinwheel centers that tend to lie over low (red) and high (blue) SF domains. In order to represent all combinations of OR and SF equally, SF was assumed to linearly vary between the two pinwheels separated by a distance d. The value of d was defined as the median minimum distance ($\pm$ mad) from one pinwheel to another whose histogram is shown in (B).}
	\label{fig:polParam}
\end{figure}

Orthogonal architectures in the PAs were not observed in our experimental data. The model we chose is based on the principle of uniform coverage that was argued as an advantageous representation of OR and SF. The ideal orthogonal architecture ensuring an optimal uniform coverage is given by a linear dependence of the SF level sets as a function of the radius $r$. Such orthogonal architectures present one extremum at the singularity. The perfect orthogonal pinwheel presenting a maximum of the SF at the singularity is given by the map: 
\begin{equation} \label{ortho}
\nu_{pol}  (r,\phi)= \nu_{max}- \beta \, r.          
\end{equation}
The parameter $\beta$ was extrapolated from the experimental data. In particular, considering a pair of pinwheels at a distance $d$ on the cortex and representing a maximum and a minimum (Fig. \ref{fig:polParam}A), the slope is identified as $\beta=\Delta \nu_{extr}/d$. We therefore estimated $d$ as the average minimal distance between PCs observed in our optical imaging data, and in particular it reads $d = 450 \pm 90 \mu m$ (Fig. \ref{fig:polParam}B). Again, we can consistently take $\nu_{max}=\Delta \nu_{extr}/2$.

Since this architecture is purely theoretical, we do not consider further deformations for the Eq. \ref{ortho}: notice that in general, adding a dependence of $\nu_{pol}  (r,\phi)$  on the angle $\phi$ would worst the precision of the detection for this kind of architecture. Changing $\nu_{max}$ into $\nu_{min}$ and $\beta$ into $-\beta$ would provide an architecture with a minimum at the PC.

Under these assumptions, the role of functional map geometry is to determine which part of the parameter space is locally represented in a given cortical area, as shown in Fig. \ref{fig:pixeldistrMain}A and Fig. \ref{fig:pixeldistrMain}B.  Notice in particular that, since the dipolar architecture is not rotationally invariant (in contrast to the orthogonal), the phase $\phi_0$ in Eq. \ref{ORmap} would reflect in a different coverage of the parameter space (compare for instance Fig. \ref{fig:pixeldistrMain}A, corresponding to $\phi_0=0$, and Fig. \ref{fig:coding}D, where $\phi_0=\pi/2$).

\subsection{Tuning Curves}

A pixel with preferred OR-SF $(\theta^*, \nu^*)$ elicits a response with amplitude $F_{\theta^* \nu^*} (\theta_{st}, \nu_{st})$ to a stimulus with OR and SF attributes $(\theta_{st}, \nu_{st})$ that achieves its maximum at $(\theta^*, \nu^*)$. For simplicity the tuning curve $F_{\theta^* \nu^*} (\theta, \nu)$ was approximated as the product of two functions $F_{\theta^*}^1$ and $F_{\nu^*}^2$
\begin{equation} \label{tuningfun}
F_{\theta^* \nu^*} (\theta, \nu) = F_{\theta^*}^1 (\theta) \, F_{\nu^*}^2 (\nu)
\end{equation}
where $F_{\theta^*}^1$ (resp. $F_{\nu^*}^2$) is the tuning curve for OR (resp. SF) detection. 
The OR tuning curve $F_{\theta^*}^1$ was modeled as a wrapped Gaussian function \cite{batschelet1981circular}:
\begin{equation}  \label{twOR}
F_{\theta^*}^1 (\theta) = {\cal N}  ~ \Sigma_{n=-N}^{N} e^{- \frac{(\theta-\theta^*+n \pi)^2}{2 \sigma_{OR}^2}} ,
\end{equation}

where $\sigma_{OR}$  is a measure of the width of the tuning curve and ${\cal N}$ is the normalisation factor. $N=3$ is sufficient to accurately describe the OR tuning response in the experimental data \cite{swindale1998orientation}. 

\begin{figure}[h]
	\centering
		\includegraphics[width=.85\textwidth]{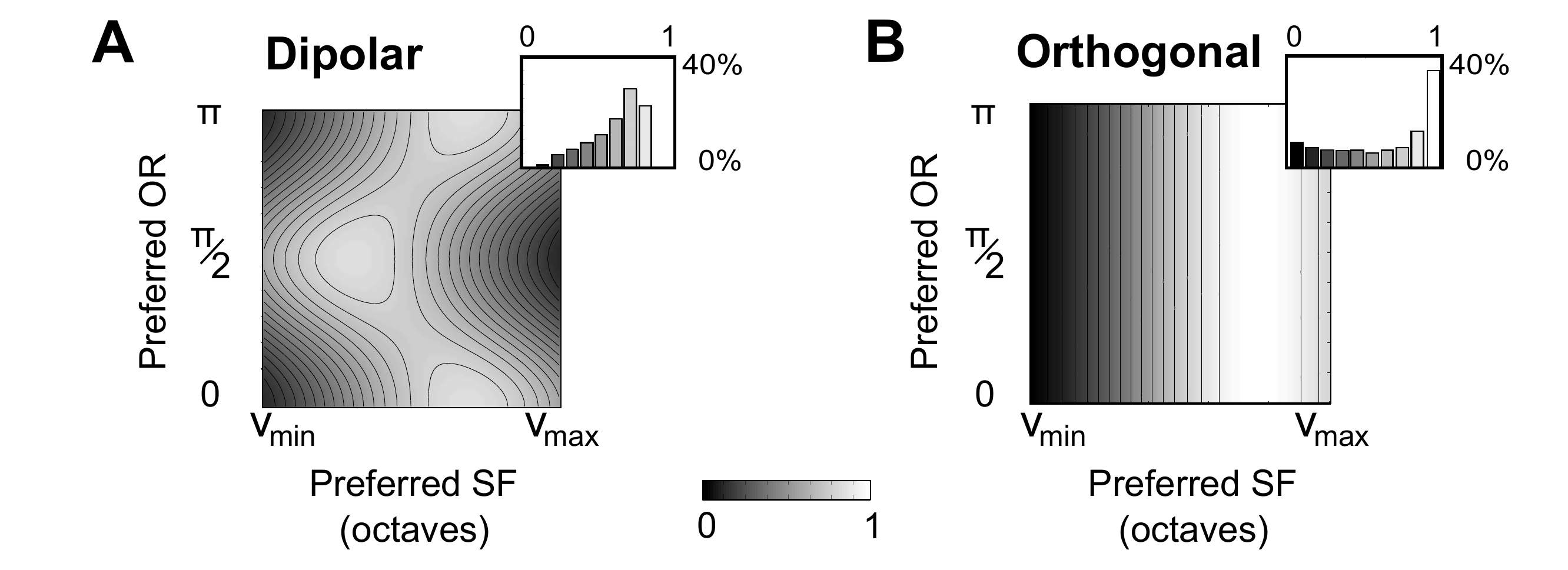}
	\caption{Relative response of dipolar (A) and orthogonal (B) architectures near PCs to external stimuli in the parameter space. For any given pixel $ (\theta^*, \nu^*)$, we show the normalised sum of all the responses $F_{\theta^* \nu^*} (\theta_{st}, \nu_{st})$ (Eq. \ref{tuningfun}, \ref{twOR} and \ref{twSF}, with parameters $\sigma_{OR}=0.63$ and $\sigma_{SF}=1$ octaves) with stimuli $(\theta_{st},\nu_{st})$ spanning the same parameter space. The normalisation is chosen by taking the minimum and maximum values of the two combined distributions. The relative distributions are represented in the histograms in the top-right corners.}
	\label{fig:pixeldistr}
\end{figure}

The SF tuning curve $F_{\nu^*}^2$ is modeled as a Gaussian function with a standard deviation $\sigma_{SF}$:
\begin{equation} \label{twSF}
F_{\nu^*}^2 (\nu)=\frac{1}{\sqrt{2 \pi} \sigma_{SF}}  e^{-\frac{(\nu-\nu^* )^2}{2 \sigma_{SF}^2}}.
\end{equation}
The Full Width at Half Height (FWHH) of the SF tuning width corresponds to $w=2 \sqrt{2 \log 2}~ \sigma_{SF}$.
The tuning width plays an important role in the coverage of the parameter space and on the coding capabilities, as we discussed in the main text. The sharper the tuning curve, the closer to Fig.~\ref{fig:pixeldistrMain} the response to pairs (OR, SF) will be, and the flatter the SF tuning curve, the more uniform the responses will be. Figure~\ref{fig:pixeldistr} show the level of response of a specific architecture to a pair (OR, SF) in the map corresponding to Fig.~\ref{fig:pixeldistrMain} of the main text. The histogram of response amplitude appears on the upper right corner. We see that in the dipolar architecture, very few pixels elicit small responses, while in the orthogonal architecture, a substantial part of the (OR, SF) plane does not elicit any response. This can probably help efficient coding and decoding. 

\subsection{The coding efficacy of dipolar and orthogonal architectures}

Once the architecture and selectivity properties are defined, we dispose of the set of responses $F_{\theta^*_i \nu^*_i} (\theta_{st}, \nu_{st})$ to a stimulus $(\theta_{st},\nu_{st})$, for all the different pixels $i$ in the cortical region under consideration. This set is used to obtain an estimation $(\theta_{ev},\nu_{ev})$ of the stimulus that elicited this response pattern in a PA. In order to extract a unique value for the estimations $(\theta_{ev},\nu_{ev})$ of the external stimulus, we took the average value of $(\theta^*_j, \nu^*_j)$ within the $10 \%$ highest amplitudes, weighted by the response amplitude $F_{\theta^*_j \nu^*_j} (\theta_{st}, \nu_{st})$. The actual mechanism used by the cortex is probably more involved, but this strategy can be seen as a reasonable first approximation of the process, or alternatively as a breech towards understanding the role of architecture in coding capabilities\footnote{Of course more complicated strategies for perception could be designed, which potentially depend on functional map topology. Designing in a relevant manner such procedure is probably premature due to the current limited understanding of perception mechanisms at the cellular level. The choice of the present coding-decoding procedure is one of the simplest assumption, and provide insight on how topology affects coding and decoding.}.
 
A set of 50 different dipolar maps (Eq. \ref{SF-def}) and 50 orthogonal (Eq. \ref{ortho}), each with parameters drawn according to the fits, were tested for 100 random stimuli $(\theta_{st},\nu_{st})$ drawn uniformly in the square $\left[0,\pi \right) \times \left[\nu_{min},\nu_{max} \right]$, with $\nu_{max} = - \nu_{min} = \Delta \nu_{extr}/2$. We first fixed the FWHH of the OR and SF tuning curves to typical values reported in the literature, i.e. respectively to $y_{exp}=80 \degree$ (near the PCs, ~\cite{rao1997opticaily}) and $w=2.48$ octaves \cite{ribot2013organization}, in order to compare the errors for dipolar and uniform coverage architectures. The errors $\epsilon_{\theta}$ and $\epsilon_{\nu}$ are defined as:
\begin{equation}\label{eq:NormalizedError}
\begin{cases}
	\epsilon_{\theta}=\left| \theta_{ev}  - \theta_{st} \right|/\pi , \\
	\epsilon_{\nu}=\left| \nu_{ev} - \nu_{st} \right|/\Delta \nu_{extr}.
\end{cases}
\end{equation}
Moreover, we repeated the same analysis for other values of the SF tuning width (see Fig. \ref{fig:tradeoff} of the main text).  

The distribution of the normalized errors is highly asymmetric. For such distributions, median and median average deviation (mad) are the relevant statistical measures. We thus obtain, for each pinwheel, a median error. This collection of 50 medians allows computing the typical errors $(\epsilon_{\theta},\epsilon_{\nu},\epsilon_{tot})$ as the median of these distributions, as well as the deviations around these values (mad).

Statistics related to differences between normalized errors were calculated using a two-sided Mann-Whitney-Wilcoxon test in Matlab\textregistered ~{\it ranksum} function. The sample size was calculated with G*Power (http://\-www.ats.ucla.edu\-/stat/gpower/), given a power equal to 0.8, an effect size of 0.8 and a probability of rejecting the null hypothesis of 0.001. 

\subsubsection{Estimating the intersection values between the error  curves} \label{sec:intersect}
The intersection point between $\epsilon_{\theta}$  and $\epsilon_{\nu}$ for the dipolar architecture in Fig. \ref{fig:tradeoff} was evaluated by fitting the error data with straight lines. To this end, we used a weighted least-squares error method for the distributions of medians described above (weights are related to the dispersion of the data). 

The linear fits for the error curves, corresponding to variations of the FWHH for the SF, named here $w$, are given by:
\begin{eqnarray}
&& \epsilon_{\theta}= 10^{-2} \left( a_{\theta} ~ w + b_{\theta}  \right),  \nonumber \\
&& \epsilon_{\nu}=10^{-2} \left( a_{\nu} ~w + b_{\nu} \right),
\end{eqnarray}  
for $1 \le w \le 4$ octaves and $a_{\theta},b_{\theta},a_{\nu}$ and $b_{\nu}$ listed in Table II, together with the correlation coefficients $|r_{\theta}|$ and $|r_{\nu}|$.
The coefficients in front of $w$ have the dimension of inverse octaves. 
The values $\overline{w}$ of the tuning width at the intersection are also listed in Table II for different choice of the parameter $\alpha$, together with the relative errors for $  \epsilon=\epsilon_{\theta}= \epsilon_{\nu}$, and are represented in the Fig. \ref{fig:alphatrend}. In particular, we show, in terms of $Z$-scores and corresponding $p$-values, that the theoretical trade-off is compatible with the experimental results $w_{exp}^{PC17} = 1.83 \pm 0.20$ octaves, when the analysis is restricted within 25 $\mu$m from PCs in A17 \footnote{Similar results are obtained when compared to the experimental value $ w_{exp}^{PC18} = 1.83 \pm 0.24$ octaves obtained in A18.}. It is also evident that the tuning width values obtained by the balance detection argument are not compatible with the value $ w_{exp}^{all}= 2.48 \pm 0.19$ octaves experimentally obtained by considering the whole V1 surface \cite{ribot2013organization} (as discussed in the main text).

\begin{table} \label{tab:cross}
  \begin{tabular}{| c || c | c | c | c | c |}
  \hline
  $\alpha$ & 0.5 &0.75 & 1 & 1.25 & 1.5 \\
  \hline
  \hline
  $a_{\theta}$ &  -1.8 $\pm$ 0.4 & -1.6 $\pm$ 0.4 & -1.5 $\pm$ 0.4 & -1.4 $\pm$ 0.4 & -1.3 $\pm$ 0.4 \\
  \hline
  $b_{\theta}$ &  9.5 $\pm$ 1.1 & 8.1 $\pm$ 1.1 & 7.3 $\pm$ 1.2 & 7.4  $\pm$ 1.1 & 6.9 $\pm$ 1.1 \\
  \hline
  $|r_{\theta}|$ &  0.99  & 0.98 & 0.98 & 0.96 & 0.96 \\
  \hline
  \hline
  $a_{\nu}$ &  5.6 $\pm$ 0.7 & 5.3 $\pm$ 0.6 & 4.8 $\pm$ 0.6 & 4.4 $\pm$ 0.5 & 4.3 $\pm$ 0.5 \\
  \hline
  $b_{\nu}$ &  -3.8 $\pm$ 1.3 & -3.6 $\pm$ 1.0 & -3.4 $\pm$ 1.0 & -3.2 $\pm$ 0.8 & -3.1 $\pm$ 0.8 \\
  \hline
   $|r_{\nu}|$ &  0.99  & 0.99 & 0.99 & 0.99 & 0.99 \\
  \hline
  \hline
  $\overline{w}$ &  1.80 $\pm$ 0.30 & 1.70 $\pm$ 0.28 & 1.73 $\pm$ 0.31 & 1.81 $\pm$ 0.30 & 1.79 $\pm$ 0.33 \\
  \hline
  $\epsilon$ &  0.06 $\pm$ 0.01 & 0.05 $\pm$ 0.01 & 0.05 $\pm$ 0.01 & 0.05 $\pm$ 0.01 & 0.04 $\pm$ 0.01 \\
  \hline
  \hline
  $Z$-score &  0.07 & 0.32 & 0.23  & 0.05  & 0.09 \\
  \hline
  $p$-value  & 0.94 & 0.75 & 0.82 & 0.96  & 0.93  \\
  \hline
  \end{tabular}
  \caption{Fits with affine functions and intersections of $\epsilon_{\theta}$ and $\epsilon_{\nu}$ curves for different values of $\alpha$, as defined in Sec. \ref{sec:intersect}. The coefficients $a_{\theta}$ and $a_{\nu}$ have the dimension of inverse octaves. The $Z$-scores and the two-sided $p$-values correspond to the null-hypothesis $\overline{w} = w_{exp}^{PC17}$, where $\overline{w}$ (expressed in octaves) is the value of the FWHH for SF at the trade-off, and $ w_{exp}^{PC17}= 1.83 \pm 0.20$ octaves (median $\pm$ mad) is the experimental result within 25 $\mu$m from PCs in A17.} 
  \end{table}

The experimental values for the SF tuning width $w$ are obtained in terms of median and mad. In order to evaluate the $Z$-scores, the robust estimate of the standard deviation (s.d.) was calculated with the formula: s.d. = 1.4826 mad. The difference of the values between the experimental data and the balanced detection point provide a value to be compared with zero, and the s.d. of the intersection point is obtained by error propagation method (valid since experimental and simulated data are statistically independent). From these data, we use the $Z$-score in two ways: (i) in order to ascertain whether two quantities are significantly distinct, we use the one-sided test, and (ii) the two-sided test in order to confirm if the two values are compatible. 

Finally notice that a correspondent trade-off can be obtained for the dipolar architecture also by varying the FWHH for OR. For example, if the SF width is fixed to $w= 2.48$ octaves, the intersection of the two $\epsilon_{\nu}$ and $\epsilon_{\theta}$ curves points for OR towards larger FWHH than the standard $80 \degree$. Once again no balance detection is observed for the orthogonal architecture.

\bibliographystyle{plain}
\bibliography{BiblioDipoles}

\end{document}